\documentclass[aps,twocolumn,showpacs]{revtex4-1}
\usepackage{graphicx}
\usepackage[all]{xy}
\usepackage{amsmath}
\usepackage{amssymb}
\usepackage{color}
\newcommand{\be}{\begin{equation}}
\newcommand{\ee}{\end{equation}}
\newcommand{\ben}{\begin{eqnarray}}
\newcommand{\een}{\end{eqnarray}}
\newcommand{\bes}{\begin{subequations}}
\newcommand{\ees}{\end{subequations}}

\begin{document}
\title{Quantum vacuum energies and Casimir forces \\ between \\ partially transparent $\delta$-function plates
}
\author{J. M. Mu$\tilde{\rm n}$oz Casta$\tilde{\rm n}$eda,$^{a}$, J. Mateos Guilarte$^b$, and A. Moreno Mosquera$^{b,c}$}
\affiliation{{\small $^a$Institut f\"{u}r Theoretische Physik, Universit\"{a}t Leipzig, Germany.\\
$^b$Departamento de F\'{\i}sica Fundamental and IUFFyM, Universidad de Salamanca, Spain.\\ $^c$Facultad Tecnologica. Universidad Distrital Francisco J. de Caldas, Colombia.}}

\date{\today}
\begin{abstract}
In this paper the quantum vacuum energies induced by massive fluctuations of one real scalar field on a configuration of two partially transparent plates are analysed. The physical properties of the infinitely thin plates are characterized by two Dirac-$\delta$ potentials. We find that an attractive/repulsive Casimir force arises between the plates when the weights of the $\delta$'s have equal/different sign. If some of the plates absorbs fluctuations below some threshold of energy (the corresponding weight is negative) there is the need to set a minimum mass to the scalar field fluctuations to preserve unitarity in the corresponding quantum field theory. Two repulsive $\delta$-interactions are compatible with massless fluctuations. The effect of Dirichlet boundary conditions at the endpoints of the interval $(-a,a)$ on a massless scalar quantum field theory defined on this interval is tantamount letting the weights
of the repulsive $\delta$-interactions to $+\infty$.
\end{abstract}

\pacs{11.27.+d, 11.25.-w}

\maketitle

\section{Introduction}
\label{sec:1}

The spectacular success of quantum field theory accurately describing scattering processes between leptons and photons in QED brought with it a striking paradox. The theoretical framework required that particles lived in open spaces of infinite volume and hence boundary conditions played no role at all {\footnote{To cope with the problems of the continuous spectrum in many QFT textbooks the system is enclosed in a \lq\lq normalization\rq\rq box of finite but very large volume. Periodic boundary conditions are imposed
on the fields in such a way that the wave vectors take values on a $3D$-lattice: $\vec{k}\in\mathbb{Z}^3$. One deals with QFT on a $3D$ torus $\mathbb{T}^3=\mathbb{S}^1\times\mathbb{S}^1\times\mathbb{S}^1$ and at the end of the calculations sends the radius of the circles to $+\infty$.}}.This fact was in sharp contrast with what happened in classical field theory where boundary conditions at the boundaries of closed (finite volume) manifolds were a central part of the theory. The situation started to change in the nineteen seventies when infrared phenomena, e.g., the quark confinement mystery, entered into the scene of quantum field theory (QFT). It was recognised very quickly, see reference \cite{Symanzik:1981wd}, that boundary conditions are implemented in QFT by surface interactions. More precisely, assuming that the quantum fields live on a 3 dimensional manifold with a surface boundary, the surface interactions are determined by adding a term proportional to a Dirac $\delta$ function to the Lagrangian on the surface times functions of the fields and their derivatives \cite{Symanzik:1981wd}.

A profound phenomenon fitting within this theoretical framework is the Casimir effect \cite{Casimir:1948dh}. In reference \cite{Bordag:1992cm} Bordag {\it et al} described analytically the two parallel plates of the ideal Casimir set-up by means of two Dirac $\delta$ potentials concentrated on a point at the centre of the plates. The Casimir energy, due to massive and massless spinor and scalar fields, was subsequently calculated setting two $\delta$-functions on the plate surfaces as electrostatic potential. These ideas were subsequently applied in the analysis of a similar effect due to field fluctuations in the presence of a penetrable spherical shell, see \cite{Bordag:1999ed}. The ultraviolet divergences appear as infinite factors multiplying the first heat trace coefficients coming from the heat kernel expansion of the free Laplacian plus a delta function potential concentrated on an infinitely thin closed surface.

More recently Fosco \textit{et al} \cite{Fosco:2009zc} derived the Casimir energy induced by the scalar field fluctuations between two finite-width mirrors by using a field derivative local expansion. The effective field theory replaces the mirrors by Dirac $\delta$ potentials, a set up that the authors reinterpret as imposing imperfect Dirichlet conditions. In reference \cite{Parashar:2012it} Milton and collaborators extend this idea to a fully electromagnetic context describing the electric permittivity and magnetic permeability in terms of $\delta$-functions with appropriate coefficients related to the plasma frequency in Barton's model for the Casimir effect on spherical shells (see ref. \cite{Barton2004}). The authors showed that the Casimir energies derived in Barton's plasma model \cite{Barton2004} are in perfect agreement with the results that they found in \cite{Parashar:2012it}. Moreover, in reference \cite{Parashar:2012it}, it was pointed out that the thin and thick boundary conditions considered by Bordag in references \cite{Bordag:2005by} and \cite{Bordag:2004dn} respectively lead to the same Casimir-Polder forces. In summary, it is suggested that the $\delta$-function interactions, admittedly an idealisation, capture the essential features of the quantum vacuum energy Casimir phenomenon. This statement is also confirmed by Fosco \textit{et al} in reference \cite{Fosco:2012gp} where the authors extend the procedures proposed in reference \cite{Fosco:2009zc} to analyse the vacuum fluctuations of the electromagnetic field disturbed by the presence of two Dirac $\delta$-plates. The boundary conditions chosen in Fosco \textit{et al} work respond to a modelization of the plate in the context of relativistic macroscopic electrodynamics. In this sense they differ from those used by Milton \textit{et al} in \cite{Parashar:2012it}, \cite{Milton:2013bm} where the boundary conditions are adapted to non-relativistic magneto-electric $\delta$-plates.

The aim of this work is to delve into the deepest analytic aspects arising from this approach to the Casimir effect by focusing our investigation on the simplest physical context. We shall thus study the quantum vacuum energy induced by the fluctuations over the real line of one real scalar massive field under the influence of two Dirac $\delta$-potentials located at two different points. Following section 2.1 of reference \cite{Milton:2004ya} we treat the quantum mechanical spectral problem posed by a double delta potential from a $(1+1)$-dimensional field theoretical perspective. Papers \cite{Graham:2003ib} and \cite{Graham:2002xq} by Jaffe and collaborators provide good evidence of interest in this task.

Specifically, we compute the Casimir energy induced by the quantum fluctuations of  one massive real scalar field between two partially transparent plates using three different methods.

\noindent $\bullet$ 1. We first analyse this problem by computing the ``reduced'' Green function and, subsequently, the $00$-component of the energy-momentum tensor. Casimir energies and forces then follow through spatial integration and ``sum'' over the frequencies. The method is identical to the procedure used in refs. \cite{Bordag:1992cm} and \cite{Milton:2004ya} and our results agree with the outcomes of these two papers. Nevertheless, we tackle a slightly more general situation. As in reference \cite{Milton:2004ya} we allow different couplings for the two Dirac $\delta$-potentials, but we also consider the possibility of having negative couplings. In this case, the $\delta$-potentials (wells) are attractive having bound states that can trap the quantum field fluctuations, i.e., we also envisage partially absorbing plates. To avoid the problems of a lack of unitarity of the quantum field theory arising in this situation we balance the mass of the fluctuations with the bound state eigenvalues in order to make fluctuation absorption impossible.

\noindent $\bullet$ 2. We apply, in a second development, the $TGTG$ prescription to calculate the Casimir energy between two compact objects from the knowledge of the ``transfer'' $T$-matrix of a single object defined by the Lipmann-Schwinger equation (see refs. \cite{Kenneth:2007jk,Bordag:2009zz}). The simplest system of two disjoint objects merely includes two points, so it is suitable to use the $TGTG$ formalism in this case to find a clarification of the structure of the Casimir energies and forces in this way. In particular, the Casimir energy of fluctuations around one single $\delta$-plate is ultraviolet divergent even after the subtraction of the contribution of quantum fluctuations around a constant background. Suitable regularisation of one-$\delta$ Casimir energy allows to show that the results obtained from the stress tensor and the $TG$-formula only differ in a finite constant. The double-$\delta$ Casimir energy, however, is finite after the subtraction of the two single-$\delta$ and empty space vacuum energies. A partial integration shows that the stress tensor result and the $TGTG$-formula provide identical results for the quantum vacuum interaction between two $\delta$-plates.

The $TGTG$ formalism offers a very good understanding of the double-$\delta$ Casimir energy as a function from the $(\alpha,\beta)$-plane of couplings (weights) into the complex plane. The real part of the Casimir energy is negative, thus producing attractive Casimir force, when both couplings have equal sign. Positive Casimir energies, hence repulsive Casimir forces, occur when the signs of the $\delta$-couplings are different. If one or both of the $\delta$-potentials, mimicking the plates, are attractive, unitarity of the quantum field theory sets a lower bound on the mass of the quantum fluctuations such that the total energy of the lowest energy state is zero. In the computation of the double-$\delta$ Casimir energy one must subtract the individual single-$\delta$'s vacuum energies. If they are attractive, the individual $\delta$-plates can absorb fluctuations of mass below the modulus of the bound state energy for a single $\delta$-potential. Therefore, allowing these light fluctuations the subtraction process renders the $TGTG$ Casimir energy of the double-$\delta$ system complex. A lower bound, equal to the sum of the two individual bound state energies on the mass of the fluctuations for the double delta system, is necessary to avoid the appearance of an imaginary part in the double-$\delta$ Casimir energy coming from the subtraction of single-$\delta$'s vacuum energies. This lower bound, however, is generically insufficient to ensure the unitarity of the quantum field theory. When both $\delta$-couplings are negative there is a small region in the plane of couplings where the lowest energy level of the double-$\delta$ system is lower than the mass lower bound. In this region the theory becomes non-unitary and a higher lower bound on the mass must be imposed to ensure the unitarity of the quantum field theory.

If the two $\delta$-interactions are repulsive there are no bound states and even the massless fluctuations give rise to a real Casimir energy and a unitary quantum field theory. In the infinitely repulsive limit of the two $\delta$-potential plates the Casimir energy induced by massless fluctuations is exactly the Casimir energy produced by two perfectly conducting plates (Dirichlet boundary conditions on the quantum fields over the plates), see references \cite{mc-asorey,jmmc-phd,Asorey:2008xt}. Thus we interpret the double-$\delta$ Casimir energy as the Casimir energy due to a family of generalised Dirichlet boundary conditions set on the quantum fluctuations of one scalar field produced by semi-transparent plates, rather than impenetrable.

\noindent $\bullet$ 3. Our third way to deal with $\delta$-potential Casimir energies is inspired by the calculation of one-loop kink mass shift (see refs. \cite{Dashen:1975hd,AlonsoIzquierdo:2011dy}). In the kink ``string'' limit of the sine-Gordon model, see ref. \cite{Mandelstam:1975hb}, the one-loop fluctuations of the scalar field in the kink background are governed by the one-$\delta$ well Hamiltonian with the threshold of the continuous spectrum displaced in such a way that the energy of the bound state is zero. Thus the Casimir energy of an attractive $\delta$-plate can be understood as the sine-Gordon kink Casimir energy in the ``string'' limit excluding the contribution of the mass renormalisation counterterm. The Casimir energy is accordingly expressed as a sum over the discrete spectrum of the square root of the eigenvalues plus the integration over the continuous spectrum of the square root of the eigenvalues weighted with the spectral density. Because the phase shifts are analytically known the calculation is completely feasible and offers a
more disclosed physical information to that encoded in either the energy-momentum tensor or the $TG$ calculation. Knowledge of the spectral data of the double $\delta$-Hamiltonian is also completely accessible allowing to explicitly write the $DHN$ formula (see ref. \cite{Dashen:1975hd}) for the two-$\delta$ Casimir energy. Bearing in mind that the double $\delta$-wells arise in the string limit of two sine-Gordon kinks it is interesting to compare this result with the quantum vacuum interaction between two sine-Gordon kinks computed by Bordag and Mu$\tilde{\rm n}$oz-Casta$\tilde{\rm n}$eda in reference \cite{Bordag:2011aa}.


\section{Quantum fluctuations of $1+1$ dimensional scalar fields}
\label{sec:2}

\subsection{The field equation and the Green function}
The fluctuations of 1D scalar fields on static classical backgrounds modelled by the function $U(x)$ are governed by the action:
\be\label{action1d}
S[\Phi]=\int\, d^2x \,\left[\frac{1}{2}\partial_\mu\Phi\partial^\mu\Phi-\frac{1}{2}U(x)\Phi^2(x,t)\right]
\ee
In order to have a well  defined scattering problem we must impose the finite area condition over the $U(x)$ classical background
\cite{galindo1990quantum}:
\be
\lim_{x\pm\infty} U(x)=m^2 \, \, , \, \, \int_{-\infty}^\infty \, dx \, (U(x)-m^2) < +\infty.
 \ee
 The classical field equation and the Green's function equation arising from (\ref{action1d}) are
\ben
 &&\left(\partial_t^2-\partial_x^2+U(x)\right)\Phi(x,t)=0\\
 && \left(\partial_t^2-\partial_x^2+U(x)\right)G(x,t;x^\prime,t^\prime)=\delta(x-x^\prime)\delta(t-t^\prime).\nonumber
\een
 Performing a Fourier decomposition in the time coordinate of the fluctuating field
 \be
 \Phi(t,x)=\int_{-\infty}^\infty \, \frac{d\omega}{2\pi}\, e^{i\omega t}\phi_\omega(x),
  \ee
and using the field equation we obtain the static fluctuation Scr\"odinger operator:
 \be
 -\phi^{\prime\prime}_\omega(x)+U(x)\phi_\omega(x)=\omega^2 \phi_\omega(x).
 \ee
The same Fourier decomposition leads to the reduced Green's function $G_\omega(x,x^\prime)$ and its corresponding differential equation:
\ben\label{red-green}
&& G(x,t;x^\prime,t^\prime)=\int_{-\infty}^\infty \, \frac{d\omega}{2\pi}\, e^{i\omega(t^\prime -t)}\, G_\omega(x,x')\\
&&\left(-\omega^2-d^2/dx^2+U(x)\right)G_\omega(x,x')=\delta(x-x').
\een
The reduced Green function plays a central role in the paper. We use the reduced Green function in the calculation of the Casimir energy using the $TGTG$ formalism developed in references \cite{Bordag:2009zz,Kenneth:2007jk,Milton:2004ya,Bordag:2011aa}.

\subsection{The stress tensor and Casimir energies}

The energy-momentum tensor arising from the action functional (\ref{action1d}) is given by
\ben
T_{\mu\nu}&=&\partial_\mu\Phi\partial_\nu\Phi-g_{\mu\nu}{\cal L}\\
{\cal L}&=&\left[\frac{1}{2}\partial_\mu\Phi\partial^\mu\Phi-\frac{1}{2}U(x)\Phi^2(x,t)\right]\nonumber
\een
The  $(0,0)$ component of the energy momentum tensor gives the energy density for any field configuration
\ben
&&T_{00}(x,t)=\frac{1}{2}\left(\left(\frac{\partial\Phi}{\partial t}\right)^2+\left(\frac{\partial\Phi}{\partial x}\right)^2+U(x)\Phi^2(x,t)\right)\nonumber\\
&&=\frac{1}{2}\left(-\Phi(x,t)\frac{\partial^2\Phi}{\partial t^2}-\Phi(x,t)\frac{\partial^2\Phi}{\partial x^2}+U(x)\Phi^2(x,t)\right). \label{pint}
\een
For those field configurations that are solutions of the equations of motion arising from (\ref{action1d}) the partial integration shown in (\ref{pint}) tells us that the energy density is given by
\begin{equation}
  T_{00}(x,t)=\left(\frac{\partial\Phi}{\partial t}\right)^2.
\end{equation}
The Green function  is also the vacuum expectation value of a time ordered product of quantum fields
\be
  G(x,t;x^\prime,t^\prime)=i\langle 0 |T(\hat{\Phi}(x,t)\hat{\Phi}(x^\prime,t^\prime))| 0 \rangle.
\ee
Therefore the vacuum expectation value of the energy density is given in terms of the Green function as \cite{Bordag:2009zz}
\be
\langle 0|\hat{T}_{00}(x)| 0 \rangle=\frac{1}{i}\left.\partial_t\partial_{t^\prime}G(x,t;x^\prime,t^\prime)\right|_{x=x^\prime, t=t^\prime}.
\ee
This is the basic formula relating the Green function to the $(0,0)$ component energy-momentum tensor. The vacuum energy is given as the integral of $\langle 0|\hat{T}_{00}(x)| 0 \rangle$ over the real line.

\subsection{The $TGTG$ method for $(1+1)$-dimensional theories}
Following references \cite{Bordag:2009zz,Kenneth:2007jk,Milton:2004ya,Bordag:2011aa} we summarise the main formulas and results that lead to the $TGTG$ formula for the Casimir energy and the quantum vacuum interaction between two compact/topological disjoint objects in $(1+1)$-dimensional scalar quantum field theories. The Lipmann-Schwinger equation arising in quantum mechanical scattering theory defines the transfer matrix \footnote{Note that the transfer matrix (name commonly used in scattering theory) is nothing else but the so called $T$ operator.} (see references \cite{2008Boya,galindo1990quantum}) as
\begin{equation}\label{l-st}
  {\bf G}= {\bf G}^{(0)}-\frac{ {\bf G}^{(0)}\cdot{\bf U}\cdot{\bf G}^{(0)}}{{\bf I}+{\bf U}\cdot{\bf G}^{(0)}}\equiv {\bf G}^{(0)}\cdot({\bf I}-{\bf T}\cdot{\bf G}^{(0)})
\end{equation}
where ${\bf G}^{(0)}$ is the Green's function for the free particle operator
\begin{equation}
{\bf K}^0=-\frac{d^2}{dx^2}+m^2 \, \,.
\end{equation}
The last equality in equation (\ref{l-st}) is written in terms of the corresponding integral kernels as
\begin{eqnarray*}
  &&G_\omega^{(U)}(x,y)=G_\omega^{(0)}(x,y)-\\&&-\int dz_1dz_2
  G_\omega^{(0)}(x,z_1)T^{(U)}_\omega(z_1,z_2)G_\omega^{(0)}(z_2,y).
\end{eqnarray*}
The general expression for the kernel of the $T$ operator can then be obtained (see reference \cite{Bordag:2011aa} for a detailed demonstration):
\begin{equation}
T_\omega(x,y)=U(x)\delta(x-y)+U(x)G_\omega^{(0)}(x,y)U(y).
\end{equation}

Compact disjoint objects in one dimension are modelled by potentials of the form
\begin{equation*}
  U(x)=U_1(x)+U_2(x),
\end{equation*}
where the smooth functions $U_i(x)$, $i=1,\,2$, have disjoint compact supports on the real line. Under this assumption the $TGTG$ formula for the vacuum interaction energy is \cite{Kenneth:2007jk}
\begin{equation}\label{tgtg-gen}
  E_0^{{\rm int}}=-\frac{i}{2}\int_0^\infty\frac{d\omega}{\pi}{\rm Tr}_{L^2}\ln\left({\bf 1}-{\bf M}_\omega\right)
\end{equation}
where the operator ${\bf M}_\omega$ and its kernel are defined as
\begin{eqnarray}
   {\bf M}_\omega&=& {\bf G}^{(0)}_\omega {\bf T}_\omega^{(1)}{\bf G}^{(0)}_\omega {\bf T}_\omega^{(2)}\label{defMop}\\
  M_\omega(x,y)&=&\nonumber{\int dz_1dz_2dz_3 \left[G^{(0)}_\omega(x,z_1)T_\omega^{(1)}(z_1,z_2)\times\right.} \\&\times&\left.G^{(0)}_\omega(z_2,z_3) T_\omega^{(2)}(z_3,y)\right]\label{kerMgen}
\end{eqnarray}
In these last expressions ${\bf T}^{(i)}_\omega$, $i=1,\,2$, is the $T$ operator associated to the object represented by $U_i(x)$, $i=1,\,2$. The potentials $U_i(x)$, $i=1,\,2$, representing the two objects define separately two Schr\"odinger problems given by the operators
\begin{equation}
{\bf K}^{(i)}_\omega=-\frac{d^2}{dx^2}+U_i(x),\quad i=1,2.
\end{equation}
In general the operators ${\bf K}^{(i)}_\omega$, $i=1,\,2$, are defined over a Hilbert space that is not isomorphic to the Hilbert space of the free quantum particle spanned by the eigenstates of the operator ${\bf K}^0$ (see \footnote{When operators ${\bf K}^{(i)}_\omega$, $i=1,\,2$ have discrete and continuum spectrum the Hilbert space spanned by eigenstates of ${\bf K}^{(i)}_\omega$, $i=1,\,2$ is not isomorphic to the one spanned by eigenstates of operator ${\bf K}^0$.}). When this happens the operator ${\bf G}^{(0)}_\omega$ is defined over a different Hilbert space than the operators ${\bf T}^{(i)}_\omega$, $i=1,\,2$. Therefore the product ${\bf G}^{(0)}_\omega\cdot{\bf T}^{(i)}_\omega$, $i=1,\,2$, is not defined and the formula (\ref{tgtg-gen}) is not valid. To avoid this problem we must perform a Wick rotation $\omega\to i\xi$: in the corresponding euclidean rotated quantum theories all the operators act over the same Hilbert space. When performing this transformation the $TGTG$ formula reads
\begin{equation}\label{tgtg-gen-eucl}
  E_0^{{\rm int}}=\frac{1}{2}\int_0^\infty\frac{d\xi}{\pi}{\rm Tr}_{L^2}\ln\left({\bf 1}-{\bf M}_{i\xi}\right).
\end{equation}
Now all the operators appearing in the $TGTG$ formula are the euclidean rotated versions of the original definition. Hence all the operators appearing act over the same Hilbert space (for more details see refs. \cite{Bordag:2011aa,Bordag:2009zz}).

Formula (\ref{tgtg-gen}) does not account explicitly for dissipative effects. However there are some dissipative phenomena that are implicit in formula (\ref{tgtg-gen}). Whenever the $L^2$ norm of the operator  ${\bf M}_{\omega}$ becomes greater than one ($\parallel{\bf M}_{\omega}\parallel_{L^2}>1$) we have that
\begin{equation}
  {\rm Tr}_{L^2}\ln\left({\bf 1}-{\bf M}_\omega\right)\in\mathbb{C}.
\end{equation}
Therefore the vacuum Casimir interaction acquires an imaginary part. Such kind of dissipative phenomena arise, for instance, when negative energy bound states appear in the one-particle states on which the quantum field theory is built. As long as unitarity is preserved all the conservation laws that are consequence of the quantum version of the N\"other theorem remain valid.

\subsection{Casimir energy from the spectral heat trace and zeta function}

The one-particle states of the quantum field theory are given by the eigenfunctions of the Schr\"odinger operator
\begin{equation}
  {\bf K}={\bf K}^0+U(x),\quad .
\end{equation}
In general this operator has both continuous  and discrete spectrum
\begin{eqnarray}
  \hspace{-0.8cm}{\bf K}\psi_j(x)&=&\omega_j^2\psi_j(x),\quad j=1,2,...,l, \,\,l\in\mathbb{N}\label{disck}\\
  \hspace{-0.8cm}{\bf K}\psi_k(x)&=&\omega(k)^2\psi_k(x;k),\,\,\omega(k)^2=k^2+m^2,\,\,k\in\mathbb{R}\label{scattk}
\end{eqnarray}
For each $k\in\mathbb{R}$ the differential equation (\ref{scattk}) has two linear independent solutions: the left-to-right incoming waves ($\psi_k^{(R)}(x)$) and right-to-left incoming waves ($\psi_k^{(L)}(x)$). The asymptotic behaviour of these solutions is determined by the scattering amplitudes (see references \citep{2008Boya,galindo1990quantum}):
\ben\label{asympscatt}
&&\psi_k^{(R)}(x)\simeq\left\{\begin{tabular}{ccc}
$e^{ikx}+r_{_R}(k)e^{-ikx}$ & $,$ & $x\to -\infty$ \\
$t(k)e^{ikx}$ & $,$ & $x\to \infty$ \\
\end{tabular}
\right.\label{asympscattR}\\
&&\psi_k^{(L)}(x)\simeq\left\{\begin{tabular}{ccc}
$t(k)e^{-ikx}$& $,$ & $x\to -\infty$ \\
$e^{-ikx}+r_{_L}(k)e^{ikx} $& $,$ & $x\to \infty$ \\
\end{tabular}
\right. .\label{asympscattL}
\een
The Wronskian of the two independent scattering solutions is given in terms of the transmission amplitude $t(k)$
\begin{equation}
  W[\psi_k^{(R)}(x),\psi_k^{(L)}(x)]=-2ik\, t(k)\equiv W_{RL}(k).
\end{equation}
The reduced Green function defined above in (\ref{red-green}) can be obtained from the two independent scattering solutions using the following expression (see \citep{Bordag:2009zz}):
\ben
G_\omega(x,x^\prime)&=&\frac{1}{W_{RL}(k)}\left(\theta(x-x^\prime)\psi_k^{(R)}(x)\psi_K^{(L)}(x^\prime)\right.\nonumber\\&&
\left.+\theta(x^\prime-x)\psi_k^{(R)}(x^\prime)\psi_K^{(L)}(x)\right),\label{red-green-scatt}
\een
where $\theta(x)$ is the Heaviside step function.
\par
Confining the whole system in a very long interval of length $L$ and imposing periodic boundary conditions over the eigenfunctions of the operator ${\bf K}$, the spectral density of the
continuous spectrum is defined as
\begin{equation}
  \varrho(k)=\frac{L}{2\pi}+\frac{d\delta(k)}{dk},
\end{equation}
where $\delta(k)=\delta_+(k)+\delta_-(k)$ is the total phase shift; the sum of the arguments of the eigenvalues of the unitary scattering matrix \citep{2008Boya}:
\begin{equation}
  e^{2 i\delta_\pm(k)}=t(k)\pm\sqrt{r_{_R}(k)r_{_L}(k)}.
\end{equation}
\par
The discrete spectrum of the operator ${\bf K}$ arises at the poles in the complex $k$-plane of the transmission amplitude $t(k)$ located in the positive imaginary momentum line $k_j=i\kappa_j$ ($\kappa_j>0$):
\be
\psi_j(x)\simeq\left\{\begin{tabular}{ccc}
$e^{\kappa_jx}$ & $,$ & $x\to -\infty$ \\
$\frac{t(i\kappa_j)}{r_{_R}(i\kappa_j})e^{-\kappa_jx}$ & $,$ & $x\to \infty$ \\
\end{tabular}
\right. .
\ee
The eigenvalues for the discrete spectrum are given by $\omega_j=-\kappa_j^2+m^2$. Note that at $k_j=i\kappa_j$ the Wronskian has poles and the doubly degeneracy of the scattering eigenfunctions disappears.

The heat trace and the spectral zeta function collect this spectral information through the definitions:
\begin{eqnarray*}
h_{\bf K}(\tau)&=&\sum_{j=1}^l\, e^ {-\tau\omega_j^2}+\int_{-\infty}^\infty \, \frac{dk}{2\pi}\, \left(L+\frac{d\delta}{dk}(k)\right)\, e^{-\tau\omega^2(k)}\\
\zeta_{\bf K}(s)&=&\sum_{j=1}^l\, \omega_j^{-2s}+\int_{-\infty}^\infty \, \frac{dk}{2\pi}\, \left(L+\frac{d\delta}{dk}(k)\right)\, \omega^{-2s}(k) \, \, , \, \,
\end{eqnarray*}
where $\tau\in\mathbb{R}^+$ and $s\in\mathbb{C}$ are respectively a positive real parameter and a complex one \footnote{We assume that there are no zero modes, $\omega^2_j\neq 0 , \forall j$, and there are no half-bound states,  $\omega_l^2<m^2$, in the spectrum of $K$. Physically, the $\tau=\frac{1}{m k_B T}$ parameter is understood as an auxiliary fictitious inverse temperature of dimensions of area: $[\tau]=L^2$.}. These two spectral functions are related by means of the Mellin transform:
\be
\zeta_{\bf K}(s)=\frac{1}{\Gamma(s)} \int_{0}^\infty \, d\tau \, \tau^{s-1} \, h_{\bf K}(\tau) \, .
\ee
Accordingly the energy induced by the quantum fluctuations around the object described by the $U(x)$ potentials measured with respect to the vacuum fluctuation energy is:
\begin{eqnarray}
E_C&=&\frac{1}{2}\left(\zeta_{\bf K}(-\frac{1}{2})-\zeta_{{\bf K}^0}(-\frac{1}{2})\right)\nonumber\\
&=&\frac{1}{2}\sum_{j=1}^l\, \omega_j-\frac{m}{4}+\frac{1}{2}\int_{-\infty}^\infty \, \frac{dk}{2\pi}\, \frac{d\delta}{dk}(k)\, \omega(k). \label{casen}
\end{eqnarray}
A subtle point  is the existence of a half-bound state of eigenvalue $\omega_{1/2}=m$ in the spectrum of ${\bf K}^0$, the constant function. According to the 1D Levinson theorem
it must be accounted for with a weight of one-half, see \cite{Barton:1984py}; this is the reason for subtracting the factor $m/4$ in formula (\ref{casen}). Another even more subtle point is the regularisation implicitly used in deriving formula (\ref{casen}). A cutoff in the energy allows us to manage a finite integration domain, $\int_{-\Lambda}^\Lambda dk$, at intermediate stages. It is known from the soliton quantisation framework that the correct regularisation, when the operator ${\bf K}$ has a finite number of bound states, is to set a cutoff in the number of fluctuation modes of both ${\bf K}$ and ${\bf K}^0$. In the limit of infinite length $L$ this prescription requires an integration by parts in the integral over the momenta, from which is obtained the quantum energy of the extended object in the form of
DHN, \cite{Dashen:1975hd}:
\begin{equation}
E_C^{{\rm DHN}}=\frac{1}{2}\sum_{j=1}^l\, \omega_j-\frac{m}{4}-\frac{1}{2}\int_{-\infty}^\infty \, \frac{dk}{2\pi}\, \frac{d\omega}{dk}(k)\, \delta(k). \label{casenDHN}
\end{equation}
For a detailed demonstration and discussion about this formula see also reference \cite{rajaraman1987solitons}.

\section{Dirac delta backgrounds. Casimir energy from the energy-momentum tensor.}
\label{sec:3}

The Dirac delta potentials have been thoroughly studied in standard textbooks on Quantum Mechanics (see for example \cite{galindo1990quantum}). In this section we introduce the results and establish the notation for the single delta and the double delta potential.
\subsection{One Dirac delta configuration background}
The potential governing massive fluctuations in the delta background is:
\be
U^{(1\delta)}(x)=m^2+\alpha\delta(x) \, \, , \, \, m^2=\mu^2+\frac{\alpha^2}{4}.
\ee
The  term $\frac{\alpha^2}{4}$ in the choice of the fluctuation mass is added to guarantee that even the possible bound state has non-negative energy. Even if we choose $\mu=0$ the vacuum energy will always be real and the corresponding quantum field theory is unitary.
The one-particle states of the quantum field theory are the eigenfunctions of the Schr\"odinger operator
\be
{\bf K}^{1\delta}={\bf K}^{0}+\alpha\delta(x) \, \, , \, \, {\bf K}^{0}=-\frac{d^2}{d x^2}+m^2\, \, .
\ee
The self-adjoint extension of the operator ${\bf K}^{1\delta}$ is defined by the Dirac delta matching conditions:
\begin{equation}
  \psi( 0^-)=\psi(0^+),\,\,\,\,\psi^\prime(0^+)-\psi^\prime(0^-)=\alpha \psi(0). \label{dmc}
\end{equation}
The continuity/discontinuity conditions (\ref{dmc}) determine subsequently the eigenstates in both the continuous and the discrete spectrum:
\begin{itemize}
\item The scattering states (continuous spectrum) for the single delta Schr\"odinger problem
\be
{\bf K}^{1\delta}\psi_k(x)=(k^2+m^2)\psi_k(x) \, \, , \, \, k\in{\mathbb R}.
\ee
are of the general form (\ref{asympscattR})-(\ref{asympscattL})  but this generic asymptotic behaviour extends to the whole real line except at the origin \cite{galindo1990quantum}:
\ben
&&\psi_k^{(R)}(x)=\nonumber{\left\{\begin{tabular}{cc}
$e^{ikx}+r(k)e^{-ikx}$ & $x<0$ \\
$t(k)e^{ikx}$  & $x>0$ \\
\end{tabular}
\right.}\\
&&\psi_k^{(L)}(x)=\nonumber{\left\{\begin{tabular}{cc}
$t(k)e^{-ikx}$ & $x<0$ \\
$e^{-ikx}+r(k)e^{ikx} $  & $x>0$ \\
\end{tabular}
\right.}.
\een
Moreover, since the Dirac delta is symmetric under parity ($\delta(-x)=\delta(x)$) the reflection amplitudes for the $R$ and the $L$ states are equal: $r_{_R}(k)=r_{_L}(k)\equiv r(k)$. Hence for a fixed $k\in\mathbb{R}$ the wave functions $\psi_k^{(R)}(x)$ and $\psi_k^{(L)}(x)$ are fully characterised by the transmission $t(k)$ and the reflection $r(k)$ amplitudes.
Solving the linear system of equations required by the Dirac delta matching conditions (\ref{dmc}) on the unknowns $t$ and $r$ one finds:
\be
t(k)=\frac{2ik}{2ik-\alpha}\, \, , \, \, r(k)=\frac{\alpha}{2ik-\alpha}.\label{scattdata-1d}
\ee
\item When the coupling is negative $\alpha<0$ and the $\delta$- potential is attractive, the scattering amplitude
 $t(k)$ has a pole in the positive imaginary axis. In this case the single delta well has a bound state:
\ben
&& k_b=-i\kappa_b=-i\frac{\alpha}{2};\,\,\,\,\, \omega_b^2=-\kappa_b^2+m^2=\mu^2\nonumber\\
&& \psi_b(x)=\sqrt{-\frac{\alpha}{2}}{\rm exp}[\frac{\alpha}{2}|x|]\, \,\nonumber .
\een
\end{itemize}
Using formula (\ref{red-green-scatt}) the reduced Green function for the single delta potential is written as
\be
 G_\omega^{(1\delta)}(x,x^\prime)=\nonumber{\left\{\begin{tabular}{cc}
$G_\omega^{(0)}-\frac{\alpha}{2ik-\alpha}
\frac{e^{ik(|x|+|x^\prime|)}}{2ik}$ & $,{\rm sgn}(x x^\prime)=+1$ \\
$-\frac{e^{ik|x-x^\prime|}}{2ik-\alpha}$  & ${\rm sgn}(xx^\prime)=-1$ \\
\end{tabular}
\right.},
\ee
whereas
\be
  G_\omega^{(0)}(x,x^\prime)= -\frac{e^{ik|x-x^\prime|}}{2ik}\nonumber
\ee
is the reduced Green function for the operator ${\bf K}^0$.

\subsection{Two Dirac deltas configuration background}
The potential governing massive fluctuations in two-delta configuration backgrounds is:
\ben
&& U^{(2\delta)}(x)=m^2+\alpha\delta(x-a)+\beta
\delta(x+a) ,\label{u2delta}  \\
&&  a>0\, \, , \, \,m^2=\mu^2+\frac{\alpha^2}{4}+\frac{\beta^2}{4}\label{m2delta} .
\een
We choose the mass of the fluctuations in such a way that the subtraction of the energy of the individual $\delta$'s will not induce spurious imaginary contributions to the vacuum energy. By doing this, even the possible bound state energies of the isolated $\delta$-wells will always be positive. This selection of the mass ensures that the corresponding quantum field theory is unitary and therefore the vacuum energy is real, except in the case that the ground state energy of the two-$\delta$ configuration is lower than the addition of the bound state energies of each $\delta$-well
separated infinitely apart. The analytical problem posed by the potential (\ref{u2delta}) is suitable for the physical description of  two parallel infinitely thin partially transparent plates.  In this
physical picture the parameters $\alpha$ and $\beta$ play the role of the plasma model frequencies mimicking the plates (see refs. \cite{Milton:2004ya,Parashar:2012it,Bordag:2007zz}).

The one-particle states of the quantum field theory are the eigenfunctions of the Schr\"odinger operator
\be
{\bf K}^{2\delta}={\bf K}^0+\alpha\delta(x-a)+\beta\delta(x+a).\label{k2delta}
\ee
The self-adjoint extension of the operator ${\bf K}^{2\delta}$ to the points $x=\pm a$ is fixed by the following Dirac delta matching conditions at these points:
\begin{eqnarray}
  &&\psi(\pm a^-)=\psi(\pm a^+),\\
  &&\psi^\prime(- a^+)-\psi^\prime(- a^-)=\alpha \psi(- a),\\
  &&\psi^\prime(a^+)-\psi^\prime( a^-)=\beta \psi( a).
\end{eqnarray}
This operator has a continuous spectrum and for certain values of the parameters $\alpha$ and $\beta$ also exhibits discrete spectrum. Both continuous and discrete spectrum eigenstates are determined from the Dirac delta matching conditions along similar lines to those described for the single delta potential:
\begin{itemize}
\item The double delta potential (\ref{u2delta}) divides the real line into three different zones: $x<-a$, $-a<x<a$ and $x>a$. Therefore, the asymptotic behaviour of the $R$ and $L$ scattering states only extends to the $x<-a$ and $x>a$ zones:
\ben
\psi_k^{(R)}(x)&=&\nonumber{\left\{\begin{tabular}{cc}
$e^{ikx}+r_{_R}(k)e^{-ikx}$ & $x<-a$ \\
$ A_{_R}(k)e^{ikx}+B_{_R}(k)e^{-ikx}$  & $-a<x<a$ \\
$t(k)e^{ikx}$ & $a<x$
\end{tabular}
\right.}\\
\psi_k^{(L)}(x)&=&\nonumber{\left\{\begin{tabular}{cc}
$t(k)e^{-ikx}$ & $x<-a$ \\
$ A_{_L}(k)e^{ikx}+B_{_L}(k)e^{-ikx}$  & $-a<x<a$ \\
$e^{-ikx}+r_{_L}(k)e^{ikx}$ & $a<x$
\end{tabular}
\right.}
\een
Note that in the intermediate zone the solutions are merely superposition of plane waves with opposite wave vector.
The Dirac delta matching conditions on these scattering waves give rise to a linear system of four algebraic equations in the four unknowns $t$, $r$, $A$, and $B$. The solution is easily obtained through Cramer's rule implemented in Mathematica:
\ben
 &&\, r_{_R}(k)=-i\frac{\alpha(2k+i\beta)e^{-2iak}+\beta(2k-i\alpha)e^{2iak}}{\Delta(k)}\label{rr2delta} \\
 && A_{_R}(k)= \frac{2k(2k+i\beta)}{ \Delta(k)}\, \, , \, \, B_{_R}(k)= -i \frac{2k\beta e^{2iak}}{ \Delta(k)}\label{ABr2delta}\\
 &&\, r_{_L}(k)=-i\frac{\alpha(2k-i\beta)e^{2iak}+\beta(2k+i\alpha)e^{-2iak}}{\Delta(k)}\label{rl2delta}\\
 && B_{_L}(k)= \frac{2k(2k+i\alpha)}{ \Delta(k)}\, \, , \, \, A_{_L}(k)= -i \frac{2k\alpha e^{2iak}}{ \Delta(k)}\label{ABl2delta} \\
 && t(k)=\frac{4 k^2}{\Delta(k)}\label{transc2delt}
\een
The common denominator $\Delta(k)$ for all the amplitudes is
\be
\Delta(k)=4 k^2+2ik(\alpha+\beta)+\left(e^{4iak}-1\right)\alpha\beta.
\ee
\item The existence of bound states in this system is determined by the poles of $t(k)$ over the positive imaginary axis in the complex $k$-plane. Note that the poles of $t(k)$ are the zeroes of the denominator $\Delta(k)$. Thus, $k=-i\kappa$ with $\kappa>0$, and these zeroes are the positive solutions of the transcendent equation
\begin{equation}\label{boundeq2delta}
e^{-4 a\kappa}= (1+\frac{2}{\alpha}\kappa)((1+\frac{2}{\beta}\kappa) .
\end{equation}
The solutions are the intersections of the  quadric $f(\kappa)=\frac{4}{\alpha\beta}\kappa^2+ \frac{2(\alpha + \beta)}{\alpha\beta}\kappa +1$ and the exponential $g(\kappa)={\rm exp}[-4 a \kappa]$. There are always two intersections for positive $\kappa$ if $|g^\prime(0)|>|f^\prime(0)|$ and only one if this inequality is not satisfied. There are two bound states if the separation between the wells ($\alpha,\beta<0$) $2a$ is such that $a> -\frac{1}{2\alpha}-\frac{1}{2\beta}=a_0$ but only one if it is shorter than this characteristic length $a_0$ of the system. The energy of the bound states becomes more and more negative with longer $a$. Figure \ref{2deltabstates} shows the distribution of the number of bound states in the $\alpha$-$\beta$ plane of couplings (see reference \cite{Guilarte:2010xn} for more details about the double delta system).
\begin{figure}[htbp]
\centerline{\includegraphics[height=7cm]{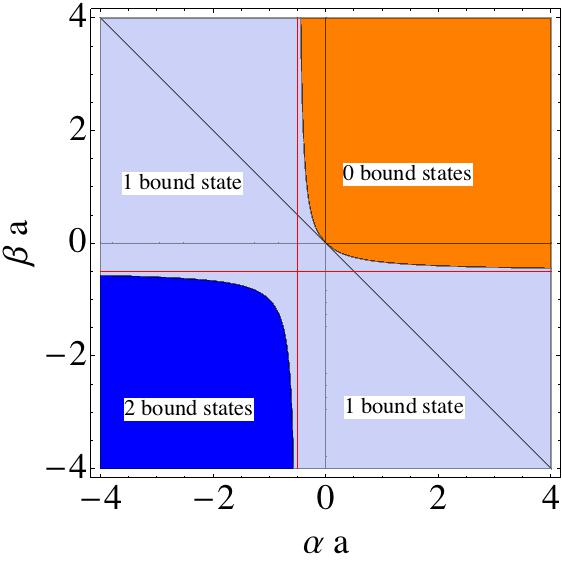}}
\caption{\footnotesize{Bound states distribution in the $(\alpha a)-(\beta a)$ plane. The different zones are limited by the branches of the hyperbola $-\frac{1}{2\alpha }-\frac{1}{2\beta }=a$ \cite{Guilarte:2010xn}.}}\label{2deltabstates}
\end{figure}
If $\kappa_j$ is a positive solution of (\ref{boundeq2delta}) the form of the bound state wave function is
\be
 \psi_j(x)=\left\{\begin{tabular}{cc}
$A(\kappa_j) e^{\kappa_j x}$ & $x<-a$ \\
$B(\kappa_j)e^{-\kappa_j x}+C(\kappa_j)e^{\kappa_j x}$  & $-a<x<a$ \\
$D(\kappa_j)e^{-\kappa_j x}$ & $a<x$
\end{tabular}
\right. \label{2dbs}  .
\ee

\end{itemize}

The two-delta reduced Green function in the zones where the two points may coincide has the structure
\begin{eqnarray*}
 G_\omega^{(2\delta)}(x,x^\prime)=G_\omega^{(0)}(x-x^\prime)+\left\{\begin{tabular}{cc}
$\delta G_{\omega,2}^{(2\delta)}(x,x^\prime)$ & $x,x^\prime<-a$ \\
$\delta G_{\omega,1}^{(2\delta)}(x,x^\prime)$  & $\vert x\vert,\vert x^\prime\vert<a$ \\
$\delta G_{\omega,3}^{(2\delta)}(x,x^\prime)$ & $x,x^\prime>a$
\end{tabular}
\right. .
\end{eqnarray*}
The different components $\delta G_{\omega,j}^{(2\delta)}(x,x^\prime)$ are computed using formula (\ref{red-green-scatt}):

\begin{eqnarray}
  \delta G_{\omega,2}^{(2\delta)}(x,x^\prime)&=&\frac{e^{-ik(x+x^\prime)}}{2k\Delta(k)}
\left[\alpha(2k+i\beta)e^{-2iak}\right.\nonumber\\
&+&\left.\beta(2k-i\alpha)e^{2iak}\right]\label{deltag2d2}\\
  \delta G_{\omega,1}^{(2\delta)}(x,x^\prime)&=& \frac{e^{2iak}}{2k\Delta(k)}\left[
\alpha(2k+i\beta)e^{ik(x+x^\prime)}\right.\nonumber\\
&+&\left.\beta(2k+i\alpha)e^{-ik(x+x^\prime)}\right]\nonumber\\
&-&\frac{i\alpha\beta}{k\Delta(k)} e^{4iak}\cos[k(x-x^\prime)]\label{deltag2d1}\\
  \delta G_{\omega,3}^{(2\delta)}(x,x^\prime)&=&\frac{e^{ik(x+x^\prime)}}{2k\Delta(k)}
\left[\alpha(2k-i\beta)e^{2iak}\right.\nonumber\\
&+&\left.\beta(2k+i\alpha)e^{-2iak}\right] .\label{deltag2d3}
\end{eqnarray}

\subsection{One-delta stress tensor and Casimir energy}
Proper zero point renormalization of the vacuum expectation value of the energy momentum tensor requires the subtraction of the vacuum expectation value of the energy momentum tensor of the free theory \citep{Bordag:2009zz}. Therefore the vacuum energy density for the scalar quantum fluctuations around a classical single delta background normalized with respect to the vacuum energy density of free field fluctuations is
\ben
&&\langle 0|\left\{\hat{T}_{00}^{1\delta}(x)-\hat{T}_{00}^{0}(x)\right\}| 0 \rangle=\nonumber\\&&\frac{1}{i}\left.\partial_t\partial_{t^\prime}\left\{G^{(1\delta)}(x,t;x^\prime,t^\prime)
-G^{(0)}(x,t;x^\prime,t^\prime)\right\}\right|_{x=x^\prime, t=t^\prime}\nonumber\\
&&=\frac{\alpha}{i}\int_{-\infty}^\infty \, \frac{d\omega}{2\pi}\, \frac{\omega^2 e^{2i\sqrt{\omega^2-m^2}|x|}}
{2i\sqrt{\omega^2-m^2}(2i\sqrt{\omega^2-m^2}-\alpha)}.
\een
To ensure that all operator products are well-defined and therefore that all integrals are well-defined it is necessary to perform an euclidean rotation $\omega=i\xi$ (see references \cite{Bordag:2009zz} and \citep{Bordag:2011aa})
\ben
&&\langle 0|\left\{\hat{T}_{00}^{1\delta}(x)-\hat{T}_{00}^{0}(x)\right\}| 0 \rangle=\nonumber\\
&&=-\frac{\alpha}{2}\int_{-\infty}^\infty \, \frac{d\xi}{2\pi}\, \frac{\xi^2 e^{-2\sqrt{\xi^2+m^2}\, \, |x|}}
{\sqrt{\xi^2+m^2}(2\sqrt{\xi^2+m^2}+\alpha)}.
\een
Using the euclidean rotated energy density the single delta Casimir energy is
\ben
&& E_C^{1\delta}(\alpha)=\int_{-\infty}^\infty \, dx \, \langle 0|\left\{\hat{T}_{00}^{1\delta}(x)-\hat{T}_{00}^{(0)}(x)\right\}| 0 \rangle=\nonumber\\
&&=-\frac{\alpha}{2}\int_{-\infty}^\infty \, \frac{d\xi}{2\pi}\, \frac{\xi^2 }
{(\xi^2+m^2)(2\sqrt{\xi^2+m^2}+\alpha)}.
\een
Performing the change of variables to $\xi^2=z^2-m^2$ as done in reference \cite{Bordag:2011aa} we obtain the stress tensor Casimir energy ($\left. E_C^{1\delta}\right\vert_{_{\rm ST}}(\alpha)$)
\be
 \left. E_C^{1\delta}\right\vert_{_{\rm ST}}(\alpha)
=-\frac{\alpha}{2\pi}\int_{m}^\infty \, dz\, \frac{\sqrt{z^2-m^2}}
{z(2z+\alpha)}.\label{inf1dcas-st}
\ee
This result was also obtained in references \cite{Milton:2004ya,Bordag:1992cm} by following the same stress
tensor procedure. Later on in this paper we will compare the Casimir energy achieved in this calculation with the outcomes for the same magnitude obtained through the use of the $T$ operator and the heat trace methods.

The integration in (\ref{inf1dcas-st}) is ultraviolet divergent, thus we choose to regularize $\left. E_C^{1\delta}\right\vert_{_{\rm ST}}$ by using an ultraviolet cutoff $\Lambda=\frac{1}{\varepsilon}$:
\ben
&&\left. E_C^{1\delta}(\alpha, \mu,\varepsilon)\right\vert_{_{ST}}= \frac{-i}{4 \pi }\sqrt{4 m^2-\alpha ^2} \times\nonumber\\
&&\times\log \left(\frac{-\alpha +i
   \sqrt{(4 m^2-\alpha ^2)(1-m^2 \epsilon ^2)}-2 m^2 \epsilon }{m
   (\alpha  \epsilon +2)}\right)\nonumber\\
 &&+\frac{1}{4 } \left(m+\sqrt{4 m^2-\alpha ^2}\right)-\frac{\alpha}{4 \pi }  \log
   \left(\sqrt{\frac{1}{m^2 \epsilon ^2}-1}+\frac{1}{m \epsilon }\right)\nonumber\\
   &&-\frac{m}{2 \pi } \arctan \left(\left(\frac{1}{m^2 \epsilon ^2}-1\right)^{-1/2}\right)\label{reg1dcas-st}
\een
To set the scale in the regulator we select the infinite mass renormalization criterion:  the finite part of the regularized Casimir energy must be zero because when $m\to\infty$ there are no massive quantum fluctuations at all !. To comply with this freezing condition of very heavy quantum fluctuations it is necessary to re-scale the regulator to be (see \footnote{Note that the new regulator $\tilde\varepsilon$ is dimensionless.}):
\begin{equation}
\varepsilon = \frac{2}{m e}\cdot \tilde{\varepsilon}
\end{equation} 
such that $\left. E_C^{1\delta}(m\to\infty)\right\vert_{_{ST,{\rm fin}}}=0$. The divergence of $\left. E_C^{1\delta}(\alpha, \mu,\tilde\varepsilon)\right\vert_{_{ST}}$ is thus regularized: $\left. E_C^{1\delta}(\alpha, \mu,\tilde\varepsilon)\right\vert_{_{ST,\,{\rm div}}}=\alpha\log(\tilde\varepsilon)/4\pi$.

\subsection{Two-delta stress tensor and Casimir energy}
The calculation of the quantum vacuum interaction energy between two delta plates requires to subtract not only the vacuum energy of the constant background but the vacuum energies of each single delta plate as well. Accordingly the renormalised two-delta Green's function that we must use to compute the quantum vacuum interaction energy is:
\begin{eqnarray*}
&&\bar{G}^{(2\delta)}_\omega(x,x^\prime;\alpha,\beta,a)= \\&&=G_\omega^{(2\delta)}(x,x^\prime;\alpha,\beta,a)-G_\omega^{(0)}(x-x^\prime)-\\
&&\hspace{-0.2cm}- G^{(1\delta)}_\omega(x+a,x^\prime+a;\alpha)-G^{(1\delta)}_\omega(x-a,x^\prime-a;\beta).
\end{eqnarray*}
The vacuum expectation value of the renormalised two-delta stress tensor $\hat{T}_{00}^{2\delta R}(x)=\hat{T}_{00}^{2\delta }(x)-\hat{T}_{00}^{1\delta }(x)-\hat{T}_{00}^{1\delta}(x)-
\hat{T}_{00}^{{\rm vac}}(x)$ is given in terms of the renormalised reduced Green function written above as
\begin{equation}
\langle 0|\hat{T}_{00}^{2\delta R}(x)|0\rangle=\int_{-\infty}^\infty \frac{dw}{2\pi i}\omega^2\bar{G}^{(2\delta)}_\omega(x,x;\alpha,\beta,a) \quad . \label{ced}
\end{equation}
Note that the integrand in this last expression for the quantum vacuum interaction energy must coincide with the spectral density:
\begin{equation}
  \frac{1}{i}\int_{-\infty}^\infty dx\,\,\bar{G}^{(2\delta)}_\omega(x,x;\alpha,\beta,a)=\frac{\varrho^{(2\delta)}(k,\alpha,\beta,a)}{k},
\end{equation}
where $k=\sqrt{\omega^2-m^2}$.
To pass from vacuum energy densities to Casimir energies we need the following integrations of the $x$-dependent functions over the three scattering zones in equation (\ref{ced}):
\begin{eqnarray*}
&& \int_{-\infty}^{-a} \, dx \, e^{-2ikx}=-\frac{e^{2iak}}{2ik}=\int^{\infty}_{a} \, dx \, e^{2ikx} \, \, , \\
&& \int_{-a}^{a} \, dx \, e^{2ikx}=\frac{{\rm sin}2a k}{k}=\int_{-a}^{a} \, dx \, e^{-2ikx} \, \, , \\ && 2\int_{-a}^{a} \, dx =4 a \quad .
\end{eqnarray*}
The integral in the third row arises from the cosine term in the Green function where both arguments lie in zone $1$. $x^\prime$ tends to $x$ either from the left or from the right.
Therefore in the modulus of the argument of the cosine both possibilities must be accounted for separately, hence the factor 2 must be included.

Using the equations (\ref{deltag2d2})-(\ref{deltag2d3}) we obtain the contribution to the spectral density from each zone:
\begin{eqnarray*}
&& \frac{1}{k} \varrho^{(2\delta)}_1(k)= \frac{i(1-e^{4iak})}{4 k^2}\left[\frac{\alpha(2k+i\beta)}{\Delta(k)}-\frac{\alpha}{2k+i\alpha}+\right.\\ &&+\left.\frac{\beta(2k+i\alpha)}{\Delta(k)}-\frac{\beta}{2k+i\beta}\right]-\frac{4ia\alpha\beta}{k\Delta(k)}e^{4iak}\\
&& \frac{1}{k}\varrho^{(2\delta)}_2(k)=\frac{i}{4k^2}\left[\frac{\alpha(2k+i\beta)}{\Delta(k)}-\frac{\alpha}{2k+i\alpha}+\right. \\ && + \left.\left(\frac{\beta(2k-i\alpha)}{\Delta(k)}-\frac{\beta}{2k+i\beta}\right)e^{4iak}\right]\\
&& \frac{1}{k}\varrho^{(2\delta)}_3(k)=\frac{i}{4k^2}\left[\left(\frac{\alpha(2k-i\beta)}{\Delta(k)}-\frac{\alpha}{2k+i\alpha}\right)e^{4iak}+\right. \\ && + \left.\frac{\beta(2k+i\alpha)}{\Delta(k)}-\frac{\beta}{2k+i\beta}\right].
\end{eqnarray*}
The total spectral density per wave number, provided by the fluctuations in the three zones, is  the sum of the quantities above, and is equal to:
\begin{eqnarray}
&&\frac{1}{k}\varrho^{(2\delta)}(k,)=\frac{1}{k} \varrho^{(2\delta)}_1(k)+\frac{1}{k} \varrho^{(2\delta)}_2(k)+\frac{1}{k} \varrho^{(2\delta)}_3(k)\nonumber\\
&&=-\frac{2\alpha\beta[1+2a(\alpha-2ik)]}{k(2k+i\alpha)\left[(2k+i\alpha)(2k+i\beta)e^{-4iak}+\alpha\beta\right]} \, \, . \label{2dwnce}
\end{eqnarray}
The Casimir energy is the integral of $\varrho^{(2\delta)}(k)/k$ weighted with $\omega^2$ over all the range of frequencies:
\begin{equation*}
E_C^{(2\delta)}(\alpha,\beta,a)=\int_{-\infty}^\infty \, \frac{d\omega}{2\pi \sqrt{\omega^2-m^2}}\, \omega^2 \, \varrho^{(2\delta)}(\sqrt{\omega^2-m^2})  .
\end{equation*}
The integral over frequencies has convergence problems due to oscillatory functions in the integrand and unitarity problems posed by bound states in Minkowski space. To avoid these problems we perform the Euclidean rotation $\omega=i\xi$ and change the variable to the imaginary momentum $\kappa=-ik=\sqrt{\xi^2+m^2}$ (see references \cite{Bordag:2009zz,Bordag:2011aa}). The Casimir energy is finally written as the integral:
\begin{eqnarray}
&&\left. E_C^{(2\delta)}(\alpha,\beta,a)\right|_{_{ST}}=\label{2dtce} \\ &&=-\frac{1}{\pi}\int_m^\infty \, d\kappa \, \frac{\alpha\beta(1+2a(\alpha+2\kappa))\sqrt{\kappa^2-m^2}}{(2\kappa+\alpha)\left[(2\kappa +\alpha)(2\kappa+\beta)e^{4a\kappa}-\alpha\beta\right]} \nonumber \, \, .
\end{eqnarray}
In the case of the double delta the integral obtained is not ultraviolet divergent because all the divergences have been subtracted by taking into account the renormalised reduced Green function $\bar{G}^{(2\delta)}_\omega(x,x^\prime;\alpha,\beta,a)$. The vacuum interaction energy between both deltas is the part of the total energy that depends on the distance between them (see \cite{Bordag:2011aa,Guilarte:2010xn,Fosco:2008vn,Bordag:1992cm,Kenneth:2007jk}).

\section{Casimir energies from the transfer matrix}
\label{sec:5}
In this section we will use the $T$ operator to compute the vacuum energy for a single semitransparent delta plate and the quantum vacuum interaction energy between two semitransparent delta plates. We start by providing the general formula for the $T$ operator generated by a potential concentrated in one point.

\subsection{The transfer matrix for potentials concentrated on points.}
The general form of the scattering states for a potential concentrated at $x=0$ is given by
\ben
&&\psi_k^{(R)}(x)=\nonumber{\left\{\begin{tabular}{cc}
$e^{ikx}+r_{_R}(k)e^{-ikx}$ & $x<0$ \\
$t_{_R}(k)e^{ikx}$  & $x>0$ \\
\end{tabular}
\right.}\\
&&\psi_k^{(L)}(x)=\nonumber{\left\{\begin{tabular}{cc}
$t_{_L}(k)e^{-ikx}$ & $x<0$ \\
$e^{-ikx}+r_{_L}(k)e^{ikx} $  & $x>0$ \\
\end{tabular}
\right.},
\een
where again $k=\sqrt{\omega^2-m^2}$.

Notice that from general scattering theory we can be sure that $t_{_R}(k)=t_{_L}(k)=t(k)$ \cite{2008Boya,galindo1990quantum}. By using formula (\ref{red-green-scatt}) the general form of the Green function for a point potential is:
\be
 G_\omega(x,x^\prime)=\nonumber{\left\{\begin{tabular}{cc}
$G_\omega^{(0)}-\frac{r_{_R}(k)}{2ik} e^{-ik(x+x^\prime)},$ & $x ,x^\prime<0$ \\
$G_\omega^{(0)}-\frac{r_{_L}(k)}{2ik} e^{ik(x+x^\prime)},$ & $x ,x^\prime>0$ \\
$G_\omega^{(0)}-\frac{t(k)-1}{2ik} e^{ik\vert x+x^\prime\vert},$  & ${\rm sgn}(xx^\prime)=-1$ \\
\end{tabular}
\right.}.
\ee
From the definition of the transfer matrix
\ben
G_\omega(x,y)\nonumber&=&G_\omega^{(0)}(x,y) \\&-&\int\, dz\, \int \, dz^\prime \, G_\xi^{(0)}(x,z)
T_\omega(z,z^\prime)G_\xi^{(0)}(z^\prime,y) \, \, \, \nonumber
\een
and using the free Green function differential equation
\be
\left(-\omega^2-\frac{d^2}{dx^2}\right)G_\omega^{(0)}(x,y)=\delta(x-y)
\ee
we get an alternative general formula for the transfer matrix
\be
T_\omega(x,y)=-\left(\omega^2+\frac{d^2}{dy^2}\right)\left(\omega^2+\frac{d^2}{dx^2}\right) \delta G_\omega(x,y),
\ee
where we defined
\be
\delta G_\omega(x,y)=\left( G_\omega(x,y) - G_\omega^{(0)}(x,y)\right)
\ee
Acting with $(\omega^2+d^2/dx^2)$ (equivalently for $(\omega^2+d^2/dy^2)$) over $\delta G_\omega(x,y)$ we always get $0$ when ${\rm sgn}(x)={\rm sgn}(y)$ because in these zones the exponentials do not contain absolute values. Hence the only non trivial contribution to the transfer matrix comes when we act with $(\omega^2+d^2/dx^2)$ and $(\omega^2+d^2/dy^2)$ over $\delta G_\omega$ in the case where ${\rm sgn}(x)\neq{\rm sgn}(y)$. The derivatives of functions that depend on absolute values are
\be
\frac{d}{d x}f(|x|)=f^\prime(|x|){\rm sgn} (x)
\ee
\be
\frac{d^2}{d x^2}f(|x|)=2f^\prime(|x|)\delta (x)+f^{\prime\prime}(|x|).
\ee
Taking into account that
\be
e^{i k|x-y|}=\nonumber{\left\{\begin{tabular}{cc}
$e^{ik(|x|+y)},$ & $x<0 ,y>0$ \\
$e^{ik(x+|y|)},$ & $x>0 ,y<0$  \\
\end{tabular}
\right.}
\ee
immediately we obtain for ${\rm sgn}(x)\neq{\rm sgn}(y)$
\be
\left(\omega^2+\frac{d^2}{dy^2}\right)\left(\omega^2+\frac{d^2}{dx^2}\right)e^{i k|x-y|}=-4k^2\delta(x)\delta(y).\nonumber
\ee
With this result the expression of the transfer matrix for arbitrary point-like potentials is given by
\be
T_\omega(x,y)=2ik (t(k)-1)\delta(x)\delta(y).\label{genT-pointV}
\ee
Note that when the potential is concentrated in another point other than zero this result is valid by simply transforming $x\mapsto x-x_0,\,\,y\mapsto y-x_0$.

\subsection{One-delta transfer matrix and Casimir energy}
Using the general formulas (\ref{genT-pointV}) and (\ref{scattdata-1d}) the euclidean rotated $T$ operator for the delta potential is
\be
T_{i\xi}^{(1\delta)}(x,x^\prime)=\frac{2\kappa\alpha}{2\kappa+\alpha}\delta(x)\delta(x^\prime) \label{matTD}\, \, \, ,
\ee
where $\kappa=\sqrt{\xi^2+m^2}$.

The $TG$ formula for the Casimir energy of the $1\delta$ configuration reads (see refs. \cite{Bordag:2011aa,Bordag:2009zz,Kenneth:2007jk})
\ben
E_C^{1\delta}(\alpha)&=&-\frac{1}{2}\int_0^\infty\frac{d\xi}{\pi}{\rm Tr}_{L^2}\ln\frac{ {\bf G}_{i\xi}^{(0)}\left({\bf 1}-{\bf T}_{i\xi}^{(1\delta)}\cdot {\bf G}_{i\xi}^{(0)}\right)}{{\bf G}_{i\xi}^{(0)}}
\nonumber \\ &=&\int_0^\infty \, \frac{d\xi}{2\pi} \, \ln \left(1-\frac{\alpha}{2\sqrt{\xi^2+m^2}+\alpha}\right)\nonumber\\
&=& \frac{1}{2\pi}\int_m^\infty \, \frac{\kappa d\kappa}{\sqrt{\kappa^2-m^2}} \, \ln\left(1-\frac{\alpha}{2\kappa+\alpha}\right).\label{TG1d}
\een
In the last part of this section we compare this result with the vacuum energy obtained from the energy momentum tensor.

\subsection{One-delta $TG$  Casimir energy ultraviolet regularization}
We regularize the vacuum energy of massive fluctuations in one-delta configuration backgrounds by
cutting the $TG$ formula at a finite ultraviolet momentum $\kappa_{\tiny u v}=\frac{1}{\varepsilon}$:
\begin{equation}
 E_C^{1\delta}(\alpha,\mu,\epsilon)=-\frac{1}{2\pi}\int_{m}^{1/\epsilon}\frac{\kappa d\kappa}{\sqrt{\kappa^2-m^2}} \, \ln \left(1-\frac{\alpha }{\alpha +2\kappa }\right) \label{uvcod} \, \, \, .
\end{equation}
The analytic integration of (\ref{uvcod}) gives, up to the leading log  approximation, the following result:
\ben
 && E_C^{1\delta}(\alpha,\mu,\epsilon)=-\frac{1}{8 \pi}\left(\alpha  \left(\ln \frac{m^2 \epsilon^2}{4} -2\right)+2\pi m\right)\nonumber\\
   &&-\frac{i \mu}{4\pi} \left( \ln \frac{(-2 \mu +i \alpha )^2}{4 m^2}-i\pi
   \right)+{\cal O}(\epsilon)\nonumber
\een
 Because $\log(x+iy)=\log(x^2+y^2)/2+i\arctan(y/x)$, we check that the regularized vacuum energy is indeed real:
\begin{eqnarray}
 && E_C^{1\delta}(\alpha,\mu,\epsilon)=\nonumber{-\frac{1}{8 \pi}\left(\alpha  \left(\ln \frac{m^2 \epsilon^2}{4} -2\right)+2\pi m\right)}\\
   &&-\frac{ \mu}{4\pi} \left(2\arctan\left(\frac{\alpha}{2\mu}\right)+\pi\right)+{\cal O}(\epsilon)\label{regvacenmas}\, \, .
\end{eqnarray}
 In the physical limit $\epsilon\to 0$  $ E_C^{1\delta}(\alpha,\mu,\epsilon)$ is logarithmic divergent. To set the scale in the regulator we select the infinite mass renormalization criterion.
 The $\mu\rightarrow\infty$ limit of (\ref{regvacenmas}) is
\begin{equation}
  E_C^{1\delta}(\alpha,\mu,\epsilon)=\frac{1}{4} \left(\frac{ \alpha}{\pi }\ln \left(\frac{2}{\mu \epsilon}\right)-2 \mu\right)+{\cal O}(\frac{1}{\mu})
\end{equation}
We thus re-scale the regulator in the form
\begin{equation}
  \epsilon=\frac{2 e^{-\frac{2 \pi  \mu }{\alpha }}}{\mu }\widetilde\epsilon
\end{equation}
in order to fit in the infinite mass renormalization criterion:
\begin{eqnarray}
  && E_C^{1\delta}(\alpha,\mu,\widetilde\epsilon)=\nonumber{-\frac{\alpha}{8 \pi}\left(\ln \frac{m^2}{\mu^2}-\frac{4 \pi  \mu }{\alpha }+ \ln
   (\widetilde\epsilon^2 )-2\right)}\\
   &&-\frac{1}{8\pi} \left(2 \pi m+2 \mu  \left(2 \arctan\left(\frac{\alpha }{2 \mu }\right)+\pi
   \right)\right)+{\cal O}(\widetilde\epsilon)\label{envacmassrf}
\end{eqnarray}
 Neglecting the logarithmic divergence, we obtain the universal finite part
\begin{eqnarray}
 && E_C^{1\delta}\left. \right|_{\rm FP}(\alpha,\mu)=\nonumber{-\frac{\alpha}{8 \pi}\left(\ln \frac{m^2}{\mu^2}-\frac{4 \pi  \mu }{\alpha }-2\right)}\\
   &&-\frac{1}{8\pi} \left(2 \pi m+2 \mu  \left(2 \arctan\left(\frac{\alpha }{2 \mu }\right)+\pi
   \right)\right)\label{envacfin}
\end{eqnarray}
that goes to zero in the $\mu\to\infty$ limit. Thus, a fine tuning of the finite renormalizations
is necessary to take into account the fact that the massive quantum fluctuations are frozen in the infinite mass limit. 

Note that with this re-scaling of the regulator, different from the re scaling used in the stress tensor version of the Casimir energy by the factor $e^{1-\frac{2 \pi  \mu }{\alpha }}$, the logarithmic divergences
of the $TG$ and $ST$ one-delta Casimir energies are the same.

\subsection{Two-delta transfer matrix and Casimir energy}

The T-matrices for the displaced delta's are immediately obtained from the $T$ operator for a single delta placed at the origin given in formula (\ref{matTD}):
\be
T^{(\alpha)}(x_1,x_2)=\delta(x_1+a)\delta(x_2+a)\cdot\frac{2\kappa\alpha}{2\kappa+\alpha}
\ee
\be
T^{(\beta)}(x_1,x_2)=\delta(x_1-a)\delta(x_2-a)\cdot\frac{2\kappa\beta}{2\kappa+\beta}
\ee
where
\begin{equation*}
  \kappa=\sqrt{\xi^2+m^2}.
\end{equation*}

 The general expression (\ref{kerMgen}) and the explicit expression for the kernels of the $T$ operator for each delta allow us to estimate the kernel of the operator ${\bf M}_\xi$ for the double-delta system:
 \begin{equation}
 M_\xi^{2\delta}(x,x^\prime)=\frac{\alpha\beta e^{-2 a \kappa}}{(2\kappa+\alpha)(2\kappa+\beta)}e^{-\kappa|x+a|}\delta(x^\prime-a).
\end{equation}
From the kernel $M_\xi^{2\delta}(x,x^\prime)$ we obtain after the corresponding integration over the whole real line the trace of the $M$-matrix in terms of $\kappa=\sqrt{\xi^2+m^2}$:
\begin{equation}
{\rm Tr}_{L^2}\, {\bf M}_\xi^{2\delta}=\int_{-\infty}^\infty \, M_\xi^{2\delta}(x,x) =\frac{\alpha\beta e^{-4\kappa a}}{(2\kappa+\alpha)(2\kappa+\beta)}.
\end{equation}

\begin{widetext}
\begin{center}
\begin{figure}[h]
\includegraphics[scale=0.33]{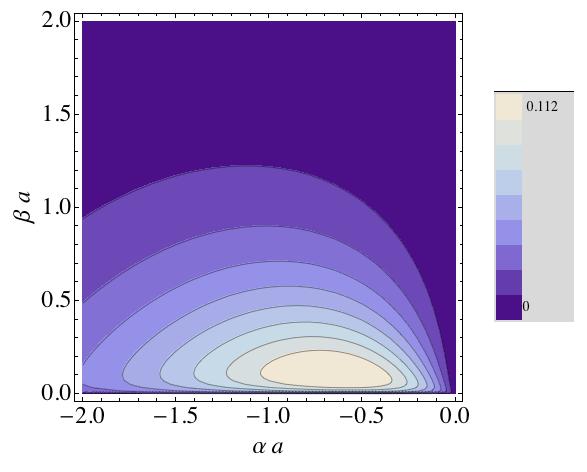}  \qquad \includegraphics[scale=0.33]{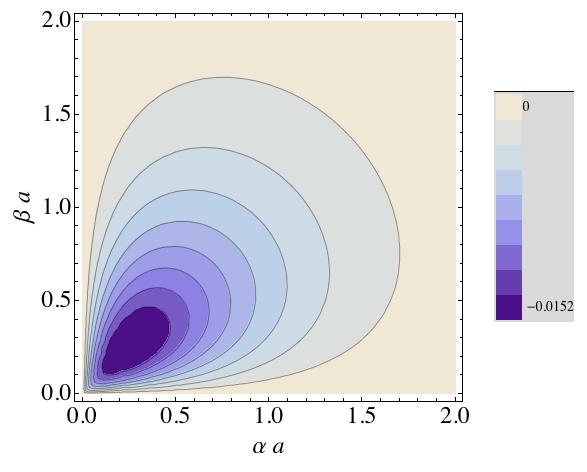}   \\  \includegraphics[scale=0.33]{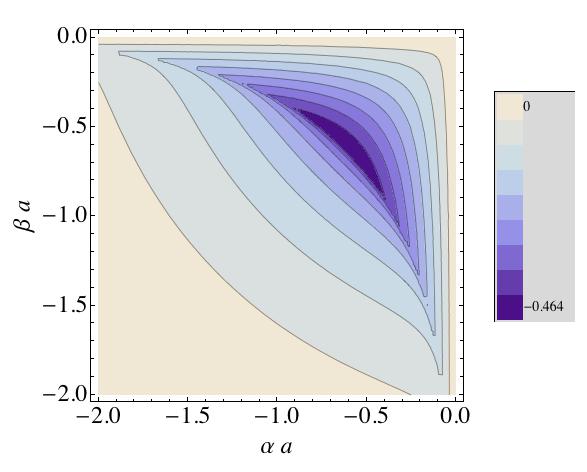}  \qquad \includegraphics[scale=0.33]{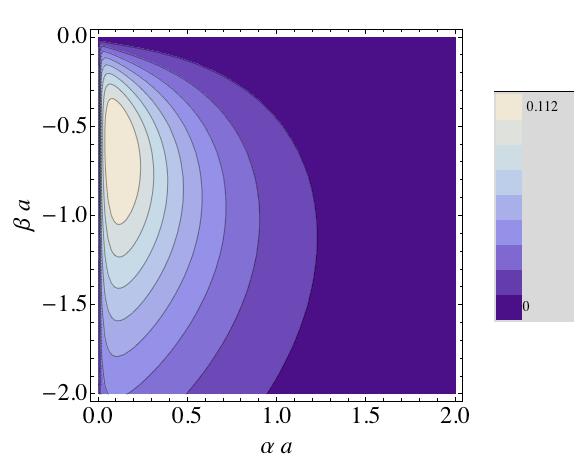}
\caption{Level curves of the real part of the non dimensional Casimir energy $2\pi a E_c^{2\delta}$ as a function of the $\alpha a$ and $\beta a$ non dimensional parameters}\label{fig2}
\end{figure}
\end{center}
\end{widetext}
The formal series expansion of $\log(1-x)$ shows that
\begin{eqnarray*}
&&{\rm Tr}_{L^2}\ln\left(1- {\bf M}^{2\delta}_\xi \right)\nonumber\\&&=\sum_{n=1}^\infty\, \frac{(-1)^{n+1}}{n}\left[{\rm Tr}_{L^2}\, {\bf M}_\xi^{2\delta}\right]^n=\ln \left(1-{\rm Tr}_{L^2}\, {\bf M}_\xi^{2\delta}\right) .
\nonumber
\end{eqnarray*}
Applying these results in the integrand of the $TGTG$ formula
\begin{equation}
 E_C^{2\delta}(\alpha,\beta,a)=\frac{1}{2}\int_0^\infty\frac{d\xi}{2\pi}{\rm Tr}_{L^2}\ln\left({\bf 1}-{\bf M}_\xi^{2\delta}\right)\label{2dtgtgc}\,
 \end{equation}
we obtain in the case of the two-$\delta$ potential the following $TGTG$ Casimir quantum vacuum energy:
\begin{equation*}
 E_C^{2\delta}(\alpha,\beta,a)=\int_0^\infty\frac{d\xi}{4\pi}
\cdot\ln\left(1-\frac{ \alpha\beta e^{-4\kappa a}}{(2\kappa+\alpha)(2\kappa+\beta)}\right)\, ,
\end{equation*}
or, more explicitly:
  \ben
&&  E_{C}^{2\delta}(\alpha,\beta, a)=\nonumber\\&=& \int_0^\infty \, \frac{d\xi}{2\pi} \, \ln\left(1-\frac{\alpha\beta e^{-4 a \sqrt{\xi^2+m^2}}}{(2\sqrt{\xi^2+m^2}+\alpha)(2\sqrt{\xi^2+m^2}+\beta)}\right)\nonumber\\&=&\frac{1}{2\pi}\int_m^\infty \frac{\kappa d\kappa}{\sqrt{\kappa^2-m^2}}
 \ln\left(1-\frac{ \alpha\beta e^{-4\kappa a}}{(2\kappa+\alpha)(2\kappa+\beta}\right)  \, \, \, , \label{2dcase}
\een
where a change of integration variable from the Euclidean energy to the Euclidean momentum has been performed in the last step. These integrals can not be carried out analytically in general. Alternatively Figure \ref{fig2} shows Mathematica plots of these integrals numerically estimated as functions of $\alpha$ and $\beta$.
\par

The results are assembled in the Figure \ref{fig2} showing selected graphs of level curves of the real part of the Casimir energy over the $\alpha a:\beta a$ plane. We observe that the Casimir energy is negative when the two $\delta$-potential plates have the same sign, they are either repulsive or attractive. If the signs are different, in the second and four quadrants, the Casimir energy, however, is positive. Therefore, the Casimir force
\begin{eqnarray*}
 && F_{C}^{2\delta}(\alpha,\beta, a)=-\frac{1}{2}\frac{d E_{C}^{2\delta}(\alpha,\beta, a)}{d a}=\\&=&-\frac{2\alpha\beta}{\pi}\int_m^\infty \frac{\kappa^2d\kappa}{\sqrt{\kappa^2-m^2}}
  \frac{1
}{e^{4 a \kappa} (2\kappa+\alpha)(2\kappa+\beta)-\alpha\beta}
\end{eqnarray*}
may be attractive or repulsive. Generically, we find attraction if the signs of the two $\delta$-potential plates are equal and repulsion when ${\rm sgn}\alpha\neq{\rm sgn}\beta$. Null Casimir energies and forces are found when one moves throughout the $\alpha:\beta$-plane of couplings. Identical qualitative behavior has been found in the Casimir interaction between two magneto-electric $\delta$-plates with dual electro-magnetic properties, see reference \cite{Milton:2013bm}. Relative minus signs in the reflection coefficients of both transverse electric and magnetic modes coming respectively from the electric permittivity and the magnetic permeabilty of the $\delta$-plate produce this interesting situation. Our scalar model will give rise to repulsive Casimir force when one of the two couplings of the $\delta$-potentials is negative. As a consequence, an imaginary part in the Casimir energy arises, unless a mass on the scalar fluctuations is introduced to avoid this dissipative phenomenon. In the case of  electromagnetic fluctuations such as those considered in \cite{Milton:2013bm} no such cutoff is needed because there are no poles in the reflection coefficients coming from bound states.

The choice $\mu=0$ in the definition of $m^2$ leaves room for an small imaginary part of the Casimir energy in a little region of the $\alpha:\beta$-plane shown in Figure \ref{fig4}.This happens because for these very weak negative values of $\alpha a$ and $\beta a$ the eigenvalue of the unique bound state of the double $\delta$-well is more negative than the sum of
the eigenvalues of the two individuals $\delta$-wells:
$-\alpha/2-\beta/2$.
\begin{widetext}
\begin{center}
\begin{figure}[h]
\includegraphics[scale=0.33]{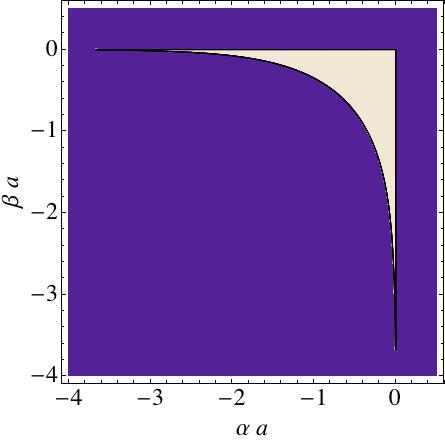} \qquad\qquad
\includegraphics[scale=0.33]{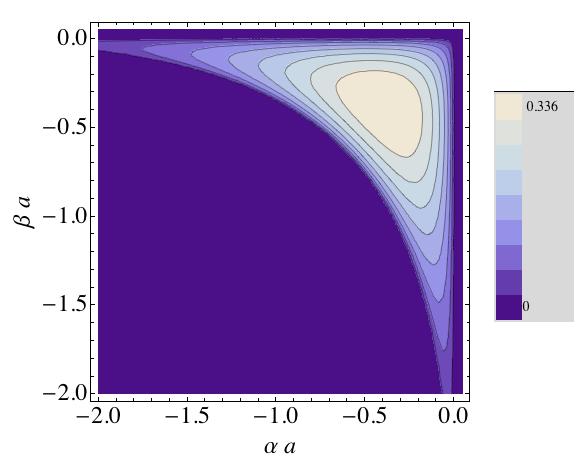}
\caption{Zone in the $(\alpha a)(\beta a)$-plane where the imaginary part of the Casimir energy is non zero and level curves of the imaginary part of the non dimensional Casimir energy $2\pi a E_c^{2\delta}$ in this region}\label{fig4}
\end{figure}
\end{center}
\end{widetext}
Thus, the $m^2=\frac{\alpha^2}{4}+\frac{\beta^2}{4}$ massive fluctuations do not push up enough the negative eigenvalue to reverse its sign and the quantum field theory is non-unitary {\footnote{Of course, there is a minimum value of $\mu>0$ such that all the eigenvalues become positive and the QFT of the system becomes unitary.}}.  When the quantum field theory is unitary the norm of the operator ${\bf M}_\xi$ is lower than one. Hence the Taylor series expansion of the logarithm that appears in the integrand of the $TGTG$ formula makes sense. Nevertheless when de Schr\"odinger problem that gives the one particle states has negative energy states the norm of ${\bf M}_\xi$ becomes bigger than one and the vacuum interaction energy has a complex value. The physical meaning of a complex vacuum energy is the surge of particle creation and annihilation in the vacuum. This effect is a bosonic cousin of the Schwinger effect where electron/positron pairs are created from the vacuum in the background of strong electric fields (see refs. \cite{Heisenberg:1935qt,Schwinger:1951nm,Dunne:2008kc}). Here, absorption and/or emission of the scalar field fluctuations by the plates is the physical phenomenon responsible of the imaginary part of the energy.

\subsubsection*{The massless Dirichlet limit: perfectly conducting plates.}
\begin{figure}
\includegraphics[scale=0.25]{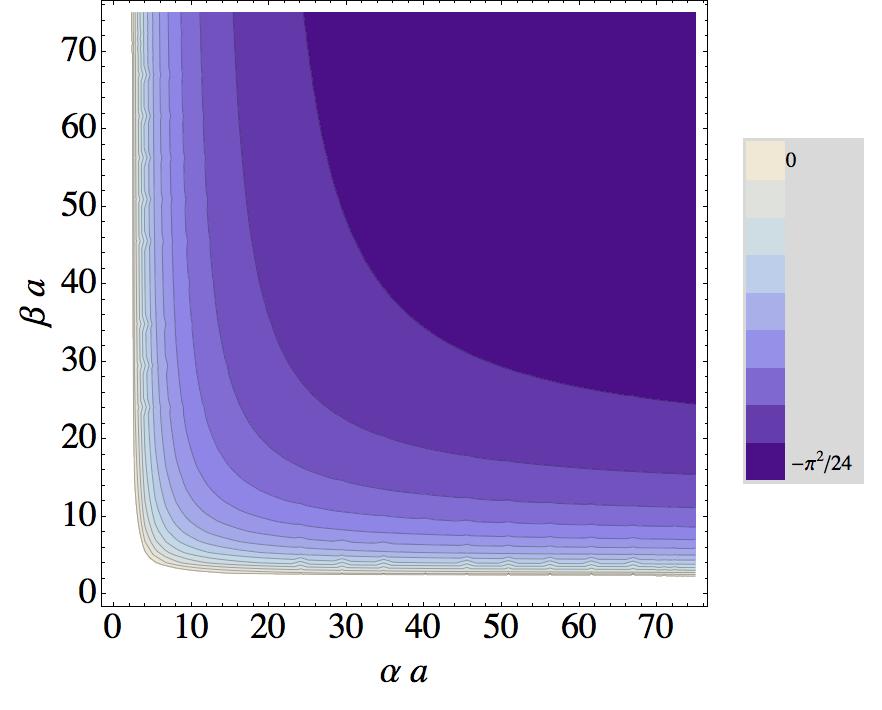}
\caption{Level curves of the non dimensional Casimir energy $2\pi a\left.E_{C}^{2\delta}(\alpha,\beta, a)\right\vert_{_{m=0}}$ in the region $\alpha,\beta>0$ }\label{fig3}
\end{figure}

When we impose $\mu=i\sqrt{\alpha^2/+\beta^2/4}$ the quantum fluctuations become massless, i. e. $m=0$. Hence from formula (\ref{2dcase}) we obtain the quantum vacuum interaction energy for massless quantum fluctuations
\begin{equation*}
\left.E_{C}^{2\delta}(\alpha,\beta, a)\right\vert_{_{m=0}}=\int_0^\infty \frac{d\kappa}{2\pi}
 \ln\left(1-\frac{ \alpha\beta e^{-4\kappa a}}{(2\kappa+\alpha)(2\kappa+\beta}\right).
\end{equation*}
Integrating by parts in this last expression we obtain an alternative formula for $\left.E_{C}^{2\delta}(\alpha,\beta, a)\right\vert_{_{m=0}}$:
\begin{eqnarray*}
\left.E_{C}^{2\delta}(\alpha,\beta, a)\right\vert_{_{m=0}}&=&-\int_0^\infty \frac{\kappa d\kappa}{2\pi}\\
 &\times &\frac{d}{d\kappa}\ln\left(1-\frac{ \alpha\beta e^{-4\kappa a}}{(2\kappa+\alpha)(2\kappa+\beta}\right)
\end{eqnarray*}
that can be directly related to the trace of operator $\left(\left.{\bf K}^{2\delta}\right\vert_{m=0}\right)^{1/2}$ by using the Cauchy's residue theorem to compute the sum over zeroes of a holomorphic function $f(z)$,
\begin{equation}
\sum_{k_n\in{\rm zeros}(f)}k_n^p=\oint_C\frac{z^p dz}{2\pi i}\frac{d}{dz}\ln\left(f(z)\right),
\end{equation}
where $C$ is a contour in the complex $z$ plane that encloses all the zeroes of $f(z)$. We stress that $\left.E_{C}^{2\delta}(\alpha,\beta, a)\right\vert_{_{m=0}}$ is only well-defined for $\alpha,\beta>0$ because the single delta vacuum energy subtraction induces an imaginary contribution when any of the delta couplings is negative. For any $\alpha,\beta>0$ the vacuum energy $\left.E_{C}^{2\delta}(\alpha,\beta, a)\right\vert_{_{m=0}}$ must be computed numerically. The level curves of $\left.E_{C}^{2\delta}(\alpha,\beta, a)\right\vert_{_{m=0}}$ as a function of $\alpha a$, and $\beta a$ can be seen in Figure \ref{fig3}.
When $\alpha,\beta\to\infty$ the integral arising in $\left.E_{C}^{2\delta}(\alpha,\beta, a)\right\vert_{_{m=0}}$ can be carried out exactly, giving rise to the very well known result of the quantum vacuum interaction energy between two perfectly conducting plates (Dirichlet boundary conditions; see reference \cite{Guilarte:2010xn}):
\begin{equation}
  \lim_{\alpha,\beta\to\infty}\left.E_{C}^{2\delta}(\alpha,\beta, a)\right\vert_{_{m=0}}=-\frac{\pi}{24\cdot(2 a)}.
\end{equation}
This is the scalar one-dimensional version of the original result obtained by H. B. G. Casimir for the electromagnetic field in the three-dimensional case in reference \cite{Casimir:1948dh}.
\subsection{Comparison between the $TG$ and stress-tensor Casimir energies}
It is worthwhile to compare the results achieved within the $TG$ formalism with the answer obtained from the stress tensor calculation. In the case of one single $\delta$ a simple partial integration makes the link between these two alternative methods:
\ben
&&E_C^{1\delta}\left|_{_{\rm TG}}\right.(\alpha)=\frac{1}{2\pi}\int_m^\infty \, \frac{z dz}{\sqrt{z^2-m^2}} \, \log
\left(1-\frac{\alpha}{2z+\alpha}\right)=\nonumber \\
&& =\frac{1}{2\pi}\sqrt{z^2-m^2}\log
\left.\left(1-\frac{\alpha}{2z+\alpha}\right)\right\vert^\infty_m \nonumber - \\ 
&& -\frac{\alpha}{2\pi}\int_{m}^\infty \, dz\, \frac{\sqrt{z^2-m^2}}
{z(2z+\alpha)}\nonumber \\ 
&& \Rightarrow E_C^{1\delta}\left|_{_{\rm TG}}\right.(\alpha)=-\frac{\alpha}{4\pi}+E_C^{1\delta}\left|_{_{\rm ST}}\right.(\alpha) \, \, \, .
\een
We see that the Casimir energies computed by these two different methods differ in a finite constant term that is precisely equal to minus $\frac{1}{2\pi}$ times the bound/anti-bound state energy.

One can also compare the Casimir energies of two $\delta$-function plates calculated within these two different procedures through partial integration in (\ref{2dcase}):
\begin{eqnarray*}
&& E_C^{2\delta}(\alpha,\beta,a)=\\ &&= \left.\frac{1}{2\pi}\sqrt{\kappa^2-m^2}\ln \left(1-\frac{\alpha\beta e^{-4 a \kappa}}{(2\kappa+\alpha)(2\kappa+\beta)}\right)\right |^\infty_m -\\
&&-\frac{\alpha\beta}{\pi}\int_m^\infty \, d\kappa \, \frac{\sqrt{\kappa^2-m^2}(1+2a(\alpha+2\kappa))}{(2\kappa+\alpha)[(2\kappa+\alpha)(2\kappa+\beta)e^{4a\kappa}-\alpha\beta]}.
\end{eqnarray*}
In contrast to what happens with a single $\delta$  the two methods lead to exactly the same result for the double delta system because the boundary term is zero:
\begin{equation}
E_C^{2\delta}(\alpha,\beta,a)\left|_{_{\rm TGTG}}\right.=E_C^{2\delta}(\alpha,\beta,a)\left|_{_{\rm ST}}\right. \quad .
\end{equation}

\section{Casimir energies from spectral functions}
\label{sec:4}
The $(1+1)$-dimensional sine-Gordon model of one scalar field is characterised by the action functional:
\begin{equation}
S[\phi]=\int \, d^2x \, \left\{\frac{1}{2}\partial_\mu\phi\partial^\mu\phi-\frac{m^4}{\lambda}\left(1-{\rm cos}\frac{\sqrt{\lambda}}{m}\phi\right)\right\} \label{sga}\, .
\end{equation}
The first-order or BPS equations in this system
\begin{equation}
\frac{d\phi}{dx}=\mp 2\frac{m^2}{\sqrt{\lambda}}{\rm sin}\frac{\sqrt{\lambda}}{2 m}\phi \label{sgbps}
\end{equation}
are solved by the kinks or solitary waves:
\begin{equation}
\phi_K(x)=\frac{m}{\sqrt{\lambda}}\left((-1)^a 4  {\rm arctan} \left(e^{\pm m(x-b)}\right)+ 2\pi n \right) \,  , \label{sGk}
\end{equation}
where $a=0,1$, $b\in\mathbb{R}$, and $n\in\mathbb{Z}$. Small deformations of these non-linear waves that are also solutions of the BPS equations (\ref{sgbps}) belong to the kernel of the first-order differential operator ${\bf\partial}=\frac{d}{dx}+m\cdot{\rm tanh}(m x)$ (we have chosen $a=b=n=0$). The second-order sine-Gordon kink fluctuations are governed by the Schr$\ddot{\rm o}$dinger operator:
\begin{equation}
{\bf K}^{\rm sG}={\bf\partial}^\dagger {\bf\partial} =-\frac{d^2}{dx^2}+m^2-\frac{2 m^2}{{\rm cosh}^2 m x} \label{sghes} \, .
\end{equation}
The spectrum of this operator is thus the basic information needed to calculate the quantum energy induced by one-loop fluctuations in the kink background. In the regime of strongly coupled and infinitely thin kink (kink string limit), $m, \lambda\to +\infty$ such that $m^2/\sqrt{\lambda}={\rm const.}\equiv\alpha/2<\infty$, we have the approximation $m^2 \cdot{\rm tanh} (m x)/\sqrt{\lambda} \simeq \, \, \frac{\alpha}{2}{\rm sgn}(x)$. In this limit the second-order operator (\ref{sghes}) becomes
\begin{equation}
{\bf K}^{1\delta}=-\frac{d^2}{dx^2}+\frac{\alpha^2}{4}-\alpha \delta(x)\, \, , \, \, \alpha>0 \label{sghesim} \, \, ,
\end{equation}
i.e., the Hamiltonian of the one-$\delta$-well shifted in such a way that the bound state energy is precisely zero. This is required by soliton physics and the reason for introducing the one-half factor in the kink string limit. Our goal in this last section is to attack the calculation of Casimir energies in $\delta$ backgrounds using the theoretical machinery developed in the study of one-loop kink fluctuations. From this point of view the quantum vacuum interaction between two Dirac deltas can be understood as the quantum vacuum interaction between two sine-Gordon kinks when both kinks can be considered infinitely thin compared with the distance between them (see reference \cite{Bordag:2011aa}).
\subsection{The one-$\delta$ Dashen-Hasslacher-Neveu formula}
We apply the Dashen-Hasslacher-Neveu formula to calculate in a third approach the Casimir energy of a $\delta$-function plate. The ingredients are spectral data of the operators
\begin{equation*}
{\bf K}^{1\delta}=-\frac{d^2}{dx^2}+m^2+\alpha \delta(x) \, \, , \, \, {\bf K}^0=-\frac{d^2}{dx^2}+m^2 \, ,
\end{equation*}
considered in preceeding sections (as before, we denote $m^2=\mu^2+\frac{\alpha^2}{4}$).
The necessary data entering in the DHN formula are collected from the discrete and continuous spectra of
${\bf K}^{1\delta}$ and ${\bf K}^0$:
\begin{enumerate}
\item Discrete spectrum.

\noindent $\bullet$ There exists a (singlet) half-bound state of ${\bf K}^0$, the constant function. Half-bound states are characterised by energies that lie in the threshold of the continuous spectrum. Thus, the ${\bf K}^0$ half-bound state eigenvalue is precisely: $\omega_m^2=m^2=\mu^2+\frac{\alpha^2}{4}$.

\noindent $\bullet$ There is a bound state in the spectrum of ${\bf K}^{1\delta}$ if and only if $\alpha<0$. The corresponding  eigenvalue is known to be: $\omega_\mu^2=m^2-\frac{\alpha^2}{4}=\mu^2$.

\item Continuous spectrum.

\noindent The information needed from the continuous spectrum is the spectral density. Choosing a very long normalisation interval of length $L$ we impose periodic boundary conditions on both the eigenfunctions of ${\bf K}^0$ and ${\bf K}^{1\delta}$:
\begin{eqnarray}
e^{iq L}&=&e^{i(k L+\delta^{1\delta}(k))}=1 \,\nonumber \\ && \Rightarrow \, q L= k L+\delta^{1\delta}(k)=2\pi n \, , \, n\in{\mathbb Z} \, \, \label{pbcsp}.
\end{eqnarray}
Here $q=\frac{2\pi}{L}n$ denotes the plane wave momenta of the ${\bf K}^0$ eigenfunctions compatible with the periodic boundary conditions. $k$, in turn, are the momenta of the scattering eigen-waves of ${\bf K}^{1\delta}$ determined from the equations on the right in (\ref{pbcsp}) in terms of the total phase shifts induced by the $one-\delta$ potential. For very large $L$ one defines the spectral densities characterising the continuous spectra to be:
\[
\varrho^0=\frac{dn}{dq}=\frac{L}{2\pi} \quad , \quad \varrho^{1\delta}=\frac{dn}{dk}=\frac{L}{2\pi}+\frac{d\delta^{1\delta}}{dk}(k)  \, \, \, .
\]

 \noindent $\bullet$ The eigenvalues in the continuous spectrum of ${\bf K}^0$ are thus $\omega^2(k)=k^2+ m^2$ whereas the spectral density is constant: $\varrho^0=\frac{L}{2\pi}$.

\noindent $\bullet$ From the phase shifts in the even and odd channels
\[
e^{2i\delta_\pm^{1\delta}(k)}=\frac{2i k \pm \alpha}{2 i k-\alpha}\quad \equiv \quad \left\{\begin{array}{c}\delta_+^{1\delta}(k)=-{\rm arctan}\frac{\alpha}{2 k} \\ \delta_-^{1\delta}(k)=0\end{array}\right.
\]
we identify the total phase shift: $\delta^{1\delta}(k)= -{\rm arctan}\frac{\alpha}{2 k}$.
The eigenvalues in the continuous spectrum of ${\bf K}^{1\delta}$ are identical to those of ${\bf K}^0$: $\omega^2(k)=k^2+ m^2$. The spectral densities, however, differ:
\[
\varrho^{1\delta}(k)=\frac{1}{2\pi}\left(L+\frac{d\delta^{1\delta}}{dk}\right) =\frac{1}{2\pi}\left(L+\frac{2\alpha}{4k^2+\alpha^2}\right)\quad .
\]
\end{enumerate}

The Casimir energy of the $\delta$-function is now provided by \lq\lq almost \rq\rq the DHN formula:
\begin{eqnarray}
E^{1\delta}_C&=& \theta(-\alpha)\frac{\mu}{2}-\frac{m}{4}+\int_{-\infty}^\infty \frac{dk}{2} \left(\varrho^{1\delta}-\varrho^0\right)\cdot\sqrt{k^2+m^2}\nonumber \\ &=& \theta(-\alpha)\frac{\mu}{2}-\frac{m}{4}\nonumber\\
&+&\int_{-\infty}^\infty \frac{dk}{2\pi}  \frac{\alpha}{4k^2+\alpha^2}
\cdot \sqrt{k^2+m^2} \label{dhnc} \quad ,
\end{eqnarray}
where $\theta(z)$ is the Heaviside step function and the additional $1/2$ factor entering in the subtraction of the half-bound state of ${\bf K}^0$ is due to the one-dimensional Levinson theorem.
The integration in (\ref{dhnc}) is ultraviolet divergent. The easiest regularisation is to use of a cutoff in the energy:
\ben
&& E^{1\delta}_C(\varepsilon) =\theta(-\alpha)\frac{\mu}{2}-\frac{m}{4}+
\frac{\alpha}{\pi}\int_0^{1/\varepsilon} \, dk \, \frac{\sqrt{k^2+m^2}}{4k^2+\alpha^2}\nonumber \\ &&= \theta(-\alpha)\frac{\mu}{2}-\frac{m}{4}+\frac{\mu}{2\pi}{\rm arctan}\left[\frac{2\mu}{\alpha}\frac{1}{\sqrt{m^2\varepsilon^2+1}}\right]+\nonumber \\ && + \frac{\alpha}{8
\pi}\log\left[\frac{(1+\sqrt{1+m^2\varepsilon^2})^2}{m^2\varepsilon^2}\right]\, \, .\label{rcasec}
\een
The \lq\lq true \rq\rq DHN formula, unveiled in quantum theory of solitons, requires a subtraction mode-by-mode to perform the vacuum zero point renormalisation because the solitonic backgrounds, in contrast to constant backgrounds, have bound states. In practical terms this procedure requires a partial integration in the contribution to the Casimir energy of the continuous part of the spectrum:
\begin{eqnarray}
\frac{1}{2\pi}\int_0^\infty \, dk \, \frac{k \delta(k)}{\sqrt{k^2+m^2}}&=&\frac{1}{2\pi}\int_0^\infty \, dk \,
\frac{\delta}{dk}(k)\sqrt{k^2+m^2}-\nonumber \\ &-& \left.\frac{1}{2\pi}\cdot \delta(k)\sqrt{k^2+m^2}\right|^\infty_0 \label{mncas} \, \, \, ,
\end{eqnarray}
such that the integral on the left-hand member of this equation (\ref{mncas}) must be used in the Casimir energy formula. Thus, the difference between taking into account all the modes up to a given energy, not equal in number in solitonic as in constant backgrounds, is determined from the phase shift at infinity and at the origin. Applied to the one-$\delta$ background the equation (\ref{mncas}) leads to the precise DHN formula for the Casimir energy:
\be
E_C^{1\delta}\left.\right|_{_{\rm DHN}}= \frac{\theta(-\alpha)}{2}(\mu-m)+\frac{\alpha}{4\pi}+\frac{\alpha}{\pi}\int_0^\infty dk  \frac{\sqrt{k^2+m^2}}{4k^2+m^2} \label{dhndce} \,  ,
\ee
i.e., only finite terms modify the result previously shown in (\ref{dhnc}) and there is no need to repeat the regularization already achieved in (\ref{rcasec}).

We mention, however, that in \cite{Toms:2012dc} a $\phi^4$ self-interactionterm has been added to the Lagrangian giving rise to a four-order vertex that induces a mass renormalization counter-term. Taking into account the contribution of this term to the Casimir energy the autor shows that the ultraviolet divergence disappears, just like in soliton physics.
\subsection{One-delta heat trace and spectral zeta function}
We now implement the spectral zeta function regularisation method to control the ultraviolet divergences in the one-$\delta$ Casimir energy. We need an intermediate tool: the associated heat trace.
\subsubsection{The ${\bf K}^{1\delta}$-heat trace}
The one-$\delta$ heat trace is defined from the spectrum of ${\bf K}^{1\delta}$ in the form:
\ben
&& h_{{\bf K}^{1\delta}}[\tau]= {\rm Tr}_{L^2}e^{-\tau K^{1\delta}} \label{htf} \\ && =\theta(-\alpha)e^{-\tau\mu^2}+\frac{1}{2\pi}\int_{-\infty}^\infty \, dk \, (L+\frac{2\alpha}{4k^2+\alpha^2})e^{-\tau(k^2+m^2)}
\nonumber \\ && =L\frac{e^{-\tau m^2}}{\sqrt{4\pi\tau}}+
\frac{e^{-\tau\mu^2}}{2}\left(1-{\rm Erf}[\frac{\alpha}{2}\sqrt{\tau}]\right)\nonumber.
\een
From the power series representation of the complementary error function
\[
{\rm Erfc}[z]=1-{\rm Erf}[z]=1-\frac{e^{-z^2}}{\sqrt{\pi}}\sum_{n=1}^\infty\, \frac{2^n}{(2n-1)!!}z^{2n-1}
\]
we infer the ${\bf K}^{1\delta}$-heat trace high-temperature expansion
\ben
\hspace{-0.5cm}h_{{\bf K}^{1\delta}}[\tau]&=&\frac{ L}{\sqrt{4\pi\tau}}e^{-\tau m^2}+\frac{e^{-\tau \mu^2}}{2}-\nonumber\\ &-&\frac{e^{-\tau m^2}}{\sqrt{4\pi}}\sum_{n=1}^\infty\, \frac{2^n}{(2n-1)!!}(\frac{\alpha}{2}\sqrt{\tau})^{2n-1} \label{asehk} \, ,
\een
In this case the series (\ref{asehk}) is truly convergent rather than asymptotic series because the integrability of the one-$\delta$ spectral problem. Previous calculations of the heat kernels coefficients
for non smooth backgrounds has been performed in reference \cite{Bordag:2004rx}. From the least equality of equation \ref{htf} and taking into account the series expansion for the error function and the exponential function the heat kernel coefficients for the delta function potential can be easily computed:
\begin{equation}
a_{-1/2}^{^{(1\delta)}}=L/\sqrt{4\pi},
\end{equation}
\begin{equation}
a_r^{^{(1\delta)}}=\frac{(-1)^r\mu^{2r}}{2\cdot r!},
\end{equation}
\begin{eqnarray}
a_{r+1/2}^{^{(1\delta)}}&=&\frac{(-1)^{r+1}}{\sqrt{4\pi}}\left[\frac{L m^{2(r+1)}}{(r+1)!}\right.\nonumber\\&+&\left.\mu^{2r}\alpha\sum_{j=0}^{r} \frac{(\alpha/2\mu)^{2j}}{j!(r-j)!\cdot (2j+1)}\right],
\end{eqnarray}
where $r=0, 1, 2,...$ is a natural number. The heat trace for the single delta system is written in terms of the heat kernel coefficients as the series:
\begin{equation}
h_{{\bf K}^{1\delta}}[\tau]=\sum_{r=0}^\infty(a_r^{^{(1\delta)}}+a_{r-1/2}^{^{(1\delta)}}\tau^{-1/2})\tau^r
\end{equation}
\subsubsection{The spectral ${\bf K}^{1\delta}$-zeta function}

The one-$\delta$ spectral zeta function $\zeta_{{\bf K}^{1\delta}}[s]$ is the Mellin's transform of the heat trace. Therefore, we obtain
\ben
&& \zeta_{{\bf K}^{1\delta}}[s]= \frac{1}{\Gamma(s)}\int_0^\infty \, d\tau \, \tau^{s-1} h_{{\bf K}^{1\delta}}[\tau] \label{melt} \\ && =
\frac{1}{2}\frac{1}{\mu^{2s}}+\frac{ L}{\sqrt{4\pi}}\frac{1}{m^{2s-1}}\frac{\Gamma(s-\frac{1}{2})}{\Gamma(s)}- \nonumber \\
&& -\frac{\alpha}{\sqrt{4\pi}}\frac{1}{m^{2s+1}}\frac{\Gamma(s+\frac{1}{2})}{\Gamma(s)}{}_2 F_1[\frac{1}{2},\frac{1}{2}+s,\frac{3}{2};-\frac{\alpha^2}{4\mu^2}] \nonumber
\een
with the spectral information encoded in the Gauss hypergeometric function ${}_2F_1[a,b,c;z]$.

The meromorphic structure of the spectral zeta function is deciphered from the Mellin's transform of the heat trace expansion:
\ben
&& \zeta_{{\bf K}^{1\delta}}[s]=\nonumber\\ &&=\frac{1}{\Gamma(s)}\int_0^\infty \, d\tau \, \tau^{s-1} \left(\frac{e^{-\tau m^2} L}{\sqrt{4\pi\tau}}+\frac{1}{2})e^{-\tau \mu^2}\right)-
\nonumber\\
&&-\frac{1}{\Gamma(s)}\int_0^\infty \, d\tau \, \tau^{s-1} \frac{e^{-\tau m^2}}{\sqrt{4 \pi}}\sum_{n=1}^\infty\, \frac{2^n}{(2n-1)!!}(\frac{\alpha}{2}\sqrt{\tau})^{2n-1}\nonumber\\
&&= \frac{ L}{\sqrt{4\pi}}\frac{1}{m^{2s-1}}\frac{\Gamma(s-\frac{1}{2})}{\Gamma(s)}+\frac{1}{2}\frac{1}{\mu^{2s}}-\nonumber\\ && -\frac{1}{\sqrt{4\pi}\alpha}\sum_{n=1}^\infty\frac{\alpha^{2n}}{2^n(2n-1)!!}\cdot
\frac{\Gamma(s+n-\frac{1}{2})}{m^{2s+2n-1}\Gamma(s)} \label{asezf}
\een
and we check that the poles of $\zeta_{{\bf K}^{1\delta}}$ arise at the points: $s+n-\frac{1}{2}=0,-1,-2, \cdots $, or, equivalently,
$s=\frac{1}{2},-\frac{1}{2},-\frac{3}{2}, -\frac{5}{2}, \cdots$.
\subsubsection{One-delta Casimir energy from the spectral zeta function}
The standard zeta function regularisation procedure prescribes a finite value for the divergent one-$\delta$
Casimir energy by assigning the result obtained from the spectral zeta function at a regular point $s\in\mathbb{C}$:
\ben
E^{1\delta}_C(s,\mu,\alpha, M)&=& \frac{1}{2}M^{2s+1}\left(\zeta_{{\bf K}^{1\delta}}[s]-\zeta_{{\bf K}^{0}}[s]\right)\label{zrce}\\
\zeta_{{\bf K}^{0}}[s]&=&2^{2s}\frac{1}{m^{2s}}+\frac{m L}{\sqrt{4\pi}}\frac{\Gamma(s-\frac{1}{2})}{m^{2s}\Gamma(s)} \nonumber \, .
\een
Here $M$ is a parameter of dimensions of inverse length introduced to keep
the dimensions of energy $L^{-1}$ at every point $s\in{\mathbb C}$. Note that we subtracted the
vacuum zero point energy also regularised by means of the corresponding spectral zeta function.

The limit $s\to -\frac{1}{2}$ where the physical Casimir energy arise, $ E^{1\delta}_C(\alpha,\mu)=\lim_{s\to -\frac{1}{2}}E^{1\delta}_C(s,\mu,\alpha, M)$, is very delicate because it is a pole of the one-$\delta$ spectral zeta function. Nevertheless, analysis of the Casimir energy near the pole allows us to isolate the divergent part:
\ben
 &&E^{1\delta}_C(\alpha,\mu)=\lim_{\varepsilon\to 0}E^{1\delta}_C(-\frac{1}{2}+\varepsilon,\mu,\alpha, M)=\nonumber\\ && =\frac{1}{4}(\mu-m)-\frac{\alpha}{8\pi}
 \lim_{\varepsilon\to 0}\frac{1}{\varepsilon}\,\cdot\, {}_2F_1[\frac{1}{2},0,
\frac{3}{2};-\frac{\alpha^2}{4\mu^2}]+\nonumber \\&& +\frac{\alpha}{8\pi}\cdot {}_2F_1^\prime[\frac{1}{2},0,
\frac{3}{2};-\frac{\alpha^2}{4\mu^2}] \, \, ,\nonumber
\een
the singularity arising at the pole of the $\Gamma$-function at the origin. To derive this formula we used:
\begin{eqnarray*}
&&\Gamma(\varepsilon)=\frac{1}{\varepsilon}\Gamma(1+\varepsilon) \, \, \, , \, \, \, {}_2F_1[\frac{1}{2},\varepsilon,
\frac{3}{2};-\frac{\alpha^2}{4\mu^2}]=\\&&={}_2F_1[\frac{1}{2},0,
\frac{3}{2};-\frac{\alpha^2}{4\mu^2}]+\varepsilon \cdot {}_2F_1^\prime[\frac{1}{2},0,
\frac{3}{2};-\frac{\alpha^2}{4\mu^2}]
\end{eqnarray*}
and the prime means derivative of the hypergeometric function with respect to the second argument.

The one-$\delta$ Casimir energy calculated from the heat kernel expansion
\ben
 && E^{1\delta}_C(\alpha,\mu)=\frac{1}{4}(\mu -m)-\\ && -\frac{1}{\alpha\sqrt{4\pi}}
 \lim_{\varepsilon\to 0}\sum_{n=1}^\infty \, \frac{\alpha^{2n}}{2^n(2n-1)!!}\frac{\Gamma(-1+\varepsilon+n)}{\Gamma(-\frac{1}{2})}\nonumber
\een
shows that the singularity only arises in the $n=1$ term.

\subsection{Two-delta Casimir energy from the spectral heat and zeta functions}
In order to compute the two-$\delta$ Casimir energies from the spectral functions we rewrite
the phase shifts of the Schr\"odinger operator $K^{2\delta}$ for two delta's of the same strength
in the form:
\ben
&& e^{2i\delta_\pm(k)}=t(k)\pm r(k)\, \, , \, \, \delta^{2\delta}(k)=\delta_+(k)+\delta_-(k)\nonumber \\ &&  e^{2i\delta_\pm(k)}=\frac{2ik+\alpha(1\pm e^{-2iak})}{2ik-\alpha(1\pm e^{2iak})} \nonumber
\een
The choice of $\delta$'s of identical weights is aimed to simplify the formulas. The spectral density on an interval of very large length, $L\to\infty$, reads:
\ben
&& \hspace{-0.6cm}\varrho^{2\delta}(k)=\nonumber \\ &&\hspace{-0.7cm}=\left(\frac{L}{2\pi}+\frac{1}{4\pi i}\frac{d}{dk}\left\{\ln\left[\frac{\alpha^2 e^{-4 i a k}-(2i k+\alpha)^2}{\alpha^2 e^{4 i a k}-(2i k-\alpha)^2}\right]\right\}\right). \label{2dspdn}
\een

\subsubsection{The ${\bf K}^{2\delta}$-spectral heat trace and zeta function}
The knowledge of the scattering data of the ${\bf K}^{2\delta}$ Schr$\ddot{\rm o}$dinger operator allows us to write the spectral heat trace and zeta function. The heat trace is
\ben
h_{{\bf K}^{2\delta}}(\tau)&=&\theta(-\alpha)e^{-\tau \omega_1^2}+\theta(-\alpha-\frac{1}{a})
e^{-\tau \omega_2^2}+\nonumber \\ &+& \int_{-\infty}^\infty\, dk \, \varrho^{2\delta}(k) \cdot e^{-\tau(k^2+m^2)}\nonumber \, \, ,
\een
formula that encodes the contribution of the (conditional) bound states (first row) and the scattering states (second row). After a partial integration we obtain:
\ben
&& h_{{\bf K}^{2\delta}}(\tau)=\label{hf2del}\\ &&=\theta(-\alpha)e^{-\tau \omega_1^2}+\theta(-\alpha-\frac{1}{a})
e^{-\tau \omega_2^2}+\frac{L}{\sqrt{4\pi\tau}}\cdot e^{-\tau m^2}+\nonumber \\ && +\tau\int_{-\infty}^\infty\, \frac{k dk}{2\pi i} \, \ln\left[\frac{\alpha^2 e^{-4 i a k}-(2i k+\alpha)^2}{\alpha^2 e^{4 i a k}-(2i k-\alpha)^2}\right] \cdot e^{-\tau(k^2+m^2)} \nonumber\, \, .
\een
The Mellin transform of these expressions leads respectively to the two-$\delta$ spectral zeta function
\ben
&&\zeta_{{\bf K}^{2\delta}}(s)=\frac{1}{\Gamma(s)}\int \, d\tau \, \tau^{s-1}\, h_{{\bf K}^{2\delta}}(\tau)= \nonumber\\ &&=
\frac{L}{\sqrt{4\pi}}\frac{\Gamma(s-\frac{1}{2})}{\Gamma(s)}\frac{1}{m^{2s-1}}+\frac{1}{\omega_1^{2s}}\theta(-\alpha)+\frac{1}{\omega_2^{2s}}\theta(-\alpha-\frac{1}{a})+\nonumber \\
&&+s\int_{-\infty}^\infty\, \frac{kdk}{2\pi i(k^2+m^2)^{s+1}}\cdot\ln\left[\frac{\alpha^2 e^{-4 i a k}-(2i k+\alpha)^2}{\alpha^2 e^{4 i a k}-(2i k-\alpha)^2}\right] \, \nonumber
\een

\subsubsection{The two body DHN formula}

Strict translation of the DHN formula would give the two-$\delta$ vacuum energy in the $s=-\frac{1}{2}$
value of the spectral zeta function as:
\be
E^{2\delta}_C\left.\right|_{DHN}=\frac{1}{2}\lim_{s\to -\frac{1}{2}}\left(\zeta_{{\bf K}^{2\delta}}(s)-\zeta_{{\bf K}^0}(s)\right)
\nonumber \, .
\ee
By doing this we fail to subtract the contributions of the self-energies of the individual $\delta$'s.
In fact the original DHN formula applies to single-body objects. To generalize this expression to two-body
structures our  criterion is to reproduce the rigorous result obtained from the $TGTG$ formula. Thus we must work with a renormalized spectral density
\ben
&&\overline{\varrho}^{2\delta}(k)=\varrho^{2\delta}(k)-\varrho^{1\delta}(k,a)-\varrho^{1\delta}(k,-a)-\varrho^0=\nonumber\\ &&=\frac{1}{4\pi i}\cdot\frac{d}{dk}\left\{\ln\left[\frac{\alpha^2 e^{-4 i a k}-(2i k+\alpha)^2}{\alpha^2 e^{4 i a k}-(2i k-\alpha)^2}\right]-\right.\nonumber\\ && \left.-\ln\left[\frac{(2 i k+\alpha)^2-\alpha^2 e^{-4 i k a}}{(2 i k-\alpha)^2-\alpha^2 e^{4 i k a}}\cdot\left(1-\frac{\alpha^2 e^{4 i k a}}{(2ik-\alpha)^2}\right)^{-1}\right] \right\}\nonumber\\ && =\frac{1}{4\pi i}\cdot\frac{d}{dk}\left\{\ln\left[1-\frac{\alpha^2 e^{4 i k a}}{(2ik-\alpha)^2}\right]\right\}\label{2dhncetgtg}
\een
where the subtraction of the individual $\delta$'s displaced with respect to each other in $2a$ exactly gives the $TGTG$ density.
From the renormalized spectral density we write the renormalized DHN two-$\delta$ Casimir energy by means of the two body DHN formula:
\ben
&& \overline{E}_C^{2\delta}\left.\right|_{\rm DHN}(\alpha, a)= \frac{\omega_1}{2}\theta(-\alpha)+\frac{\omega_2}{2}\theta(-\alpha-\frac{1}{a})-\frac{m}{4}-\nonumber \\
&&-\mu\theta(-\alpha)-\frac{1}{4 \pi i}\int_0^\infty \, \frac{kdk}{\sqrt{k^2+m^2}}\ln\left[1-\frac{\alpha^2 e^{4 i k a}}{(2i k-\alpha)^2}\right] \nonumber \, .
\een
Note that we have subtracted also the square roots of the eigenvalues of the half-bound state of the free Hamiltonian and the two bound state eigenvalues of the individual $\delta$'s.

\section{Prospects and further Comments}
\subsection{Vacuum energy of many $\delta$-interactions}
The concepts explored and the techniques developed in the core of this paper can be extended to analyse an array of $2N+1$ Dirac $\delta$-potentials. Here $N\in\mathbb{N}^*$ or $N\in\mathbb{N}_2^*$, i.e., $N$ is a positive integer or half-integer in such a way that $2 N+1$ is either an odd or even integer. If $n=-N, -N+1, \cdots , N-1, N$ span the integers or half-integers between $-N$ and $N$ we consider the background
\begin{equation}\label{Nbac}
U(x)=\sum_{n=-N}^N\, \alpha_n \, \delta(x-2 n a)  \, \, \, ,
\end{equation}
e.g., for $N=0$ it gives the one-$\delta$-potential, the $N=\frac{1}{2}$ case corresponds to two-$\delta$'s, $N=1$ to three $\delta$'s, etc. The quantum vacuum energy induced by the fluctuations of an scalar field on this background (\ref{Nbac}) is computable by plugging the kernel of the ${\bf M}$-operator
\begin{eqnarray*}
&& M_\omega^{(\alpha_{-N}, \alpha_{-N+1} \cdots ,\alpha_{N-1}, \alpha_N)}(x_1,x_2)\\&=&\int dz_1dz_2 \cdots dz_{4N+1}\left[ G_\omega^{(0)}(x_1,z_1)T_\omega^{(\alpha_{-N})}(z_1,z_2)\right. \cdot \\ && \cdot G_\omega^{(0)}(z_2,z_3)T_\omega^{(\alpha_{-N+1})}(z_3,z_4)\cdots \\ && \cdots \left. G_\omega^{(0)}(z_{4 N},z_{4 N+1})T_\omega^{(\alpha_{N})}(z_{4 N+1},x_2)\right]
\end{eqnarray*}
in formula (\ref{tgtg-gen}). Recall that in the Euclidean version we have
\be
G_{i\xi}^{(0)}(x_1,x_2)=\frac{1}{2\kappa}\cdot e^{-\kappa\vert x_1-x_2\vert} \nonumber
\ee
\be
T_{i\xi}^{(\alpha_n)}(x_1,x_2)=\delta(x_1-2 n a)\delta(x_2-2 n a)\cdot
\frac{2\kappa\alpha_n}{2\kappa+\alpha_n}\nonumber\, \,
\ee
formulas from which we derive
\ben
&& M_{i\xi}^{(\alpha_{-N}, \cdots , \alpha_N)}(x_1,x_2)=\nonumber \\ && = e^{-\kappa\vert x_1+2 N a\vert}\delta(x_2-2 N a)\cdot\prod_{n=-N}^N\frac{\alpha_n}{2\kappa+\alpha_n}\cdot e^{-2\vert n\vert \kappa a}
\nonumber
\een
and
\ben
&& {\rm Tr}_{L^2} {\bf M}_{i\xi}^{(\alpha_{-N}, \cdots , \alpha_N)}=\int_{-\infty}^\infty dx
M_{i\xi}^{(\alpha_{-N}, \cdots , \alpha_N)}(x,x)=\nonumber\\ &&= e^{-4 N \kappa a}\cdot \prod_{n=-N}^N\frac{\alpha_n}{2\kappa+\alpha_n}\cdot e^{-2\vert n\vert \kappa a} \label{Ntdel} \, .
\een
The Lipmann-Schwinger equation determining the transfer matrix provides us with a result for the vacuum energy of an array of $2 N+1$-$\delta$'s where the vacuum energies of any array with a lower number of
$\delta $'s (including the constant background) are subtracted. Thus, the $TG$ procedure applied to $2 N+1$
$\delta $'s not only subtracts the divergent vacuum zero point energy and the $2 N+1$ Casimir \lq\lq self-energies \rq\rq of the individual $\delta$'s but also the finite two-body, three-body, $\cdots$ , 2$N$-body
Casimir energies. In the recent reference \cite{Shajesh:2011ef}, however, the interactions between a lower number of $\delta$'s than $N$ have been also considered.

\subsection{Supersymmetric $\delta$-interactions}

The fluctuations of two real scalar fields on two possibly different static bacgrounds are governed by the action:
\ben
&& S[\Phi_+,\Phi_-]=\nonumber\\&=&\frac{1}{2}\int\, d^2x \, \left\{ \partial_\mu\Phi_+\partial^\mu\Phi_+ +(m_+^2+U_+(x))\Phi_+^2(t,x) +\nonumber \right.\\ &+& \left.\partial_\mu\Phi_-\partial^\mu\Phi_- +(m_-^2+U_-(x))
\Phi_-^2(t,x) \right\} \label{2scff} \, \, .
\een
The one-particle states of the quantum field theories are obtained through a Fourier transform from time
to frequency of the eigenfunctions of the Schr$\ddot{\rm o}$dinger operators:
\be
{\bf K}_\pm=-\frac{d^2}{dx^2}+U_\pm(x) \nonumber .
\ee
Generically, this system is no more than the system considered in the main core of this paper counted twice.
There is, however, an interesting situation: if $m_+^2=m_-^2=m^2$ it may be possible that $K_+$ and
$K_-$ are supersymmetric partners and their spectra have the symmetry properties arising in supersymmetric
quantum mechanics. Briefly, the structure is the following:
\begin{enumerate}
\item One starts from the \lq\lq supercharges \rq\rq
\be
{\bf Q}=\left(\begin{array}{cc}0 & \frac{d}{dx}+W(x)\\ 0 & 0 \end{array}\right) \, \, , \, \, {\bf Q}^\dagger=\left(\begin{array}{cc}0 & 0 \\ -\frac{d}{dx}+W(x) & 0 \end{array}\right)\nonumber
\ee
where $W(x)$ is a real function called the superpotential.
\item The supersymmetric Hamiltonian is:
\ben
&&{\bf K}= \frac{1}{2}\left({\bf Q}{\bf Q}^\dagger+{\bf Q}^\dagger{\bf Q}\right)= \nonumber \\ &=&\frac{1}{2}\left(\begin{array}{cc} -\frac{d^2}{dx^2}+\frac{dW}{dx}\frac{dW}{dx}+\frac{d^2W}{dx^2} & 0 \\ 0 & -\frac{d^2}{dx^2}+\frac{dW}{dx}\frac{dW}{dx}-\frac{d^2W}{dx^2}\end{array}\right)\nonumber \, .
\een
\item The supersymmetry algebra
\be
\{ {\bf Q},{\bf Q}^\dagger\}=2{\bf H} \, \, \, , \, \, \, [{\bf H},{\bf Q}]=[{\bf H},{\bf Q}^\dagger]=0 \nonumber
\ee
shows that the the supercharges are \lq\lq super \rq\rq symmetry operators.
\end{enumerate}
The two kinds of scalar fluctuations give rise to a supersymmetric quantum mechanical problem if and only if:
\be
U_\pm(x)=\frac{dW}{dx}\frac{dW}{dx}\pm \frac{d^2W}{dx^2} \nonumber \, \, .
\ee
In the case of two $\delta$-function plates supporting the fluctuations of two scalar fields one spectral
problem in supersymmetric quantum mechanics arises if the backgrounds are of the form:
\be
U_\pm(x)=\pm \alpha \delta(x+a) \pm \beta \delta(x-a)+\alpha^2\theta(-x-a)+\beta^2\theta(x-a) \nonumber \, .
\ee
The hidden reason for supersymmetry is the existence of a superpotential
\be
W(x)=\frac{\alpha}{2}\cdot\vert x+a\vert + \frac{\beta}{2} \cdot\vert x-a\vert + \frac{\beta-\alpha}{2}\cdot x \nonumber
\ee
that defines the supercharges of this problem. Standard lore in supersymmetric quantum mechanics, see e.g. \cite{Wipf:2005sk} and references quoted therein, classifies the characteristics of the spectrum of a supersymmetric Hamiltonian as follows:
\begin{itemize}

\item The ground state of ${\bf K}$ may be unique or doubly degenerate. If it is unique the ground state
belongs to the spectrum of either ${\bf K}_+$ or ${\bf K}_-$. In this case there exists spectral asymmetry, and supersymmetry
is unbroken. Degenerate ground states form a doublet: one member belongs to the spectrum of ${\bf K}_+$ and the
other is an eigenstate of ${\bf K}_-$, and supersymmetry is spontaneously broken. The two-$\delta$ supersymmetric Hamiltonian exhibits a unique ground state and supersymmetry is unbroken.

\item Higher excited levels in the discrete spectrum of ${\bf K}$ come in degenerate pairs living
respectively in the spectrum of ${\bf K}_+$ and ${\bf K}_-$. Their positive eigenvalues are identical. Hence, operators ${\bf K}_+$ and ${\bf K}_-$ are almost isospectral.

\item The continuous spectrum eigenvalues of ${\bf K}_+$ and ${\bf K}_-$ are also positive and identical.
The spectral densities of the SUSY pair of operators, however, are different but are related by the supercharges.
\end{itemize}
In summary, once we know the Casimir energy due to one kind of fluctuations the contribution of the other kind follows from supersymmetry. A warning: in order to apply the $TGTG$ formula in this supersymmetric context
we must rely on the Green's function and the $T$-matrix in one step/delta background, e.g., located at the origin:
\be
U_{\rm s\delta}(x;\alpha,s)=\alpha \delta(x)+s^2(1-\theta(x)) \label{stepd} \, \, .
\ee
In this situation (no time reversal invariance) the transmission amplitudes are also different:
\ben
&&t_R(\omega)=\frac{2q}{\Delta_{\rm s\delta}} \, \, ,
\, \, t_L(\omega)=\frac{2k}{\Delta_{\rm s\delta}} \, \, ; \, \,\\
&&r_R(\omega)=\frac{q-k-i \alpha}{\Delta_{\rm s\delta}} \, \, \, , \, \, \, r_L(\omega)=\frac{k-q-i \alpha}{\Delta_{\rm s\delta}}
\een
\be
\Delta_{\rm s\delta}(\omega,\alpha)=k+q+i\alpha \, \, \, , \, \, \,\omega^2=k^2=q^2+s^2  ,
\ee
\be
W(\psi_\omega^{(R)},\psi_\omega^{(L)})= \frac{4ikq}{\Delta_{\rm s\delta}}
\ee
where $k$ and $q$ are the momenta of the plane waves on the $x>0$ and $x<0$ half-lines respectively. From the Wronskian and the scattering waves incoming from the left or the right we obtain the reduced Green's function and the kernel of the ${\bf T}$-operator in four pieces:
\begin{widetext}
\be
G_\omega^{(0)}(x,y)=\left\{\begin{array}{ccc} -\frac{1}{2 i k}\left(e^{ik\vert x-y\vert}-(\frac{2 k}{\Delta_{\rm s\delta}}-1)e^{ik(x+y)}\right) & , \, x>0 & , \, y>0 \\ \\ i\frac{1}{\Delta_{\rm s\delta}}e^{i(kx-qy)} & , \, x>0 & , \, y<0 \\ \\ i\frac{1}{\Delta_{\rm s\delta}}e^{i(ky-qx)} & , \, x<0 & , \, y>0 \\ \\ -\frac{1}{2 i q}\left(e^{iq\vert x-y\vert}-(\frac{2 q}{\Delta_{\rm s\delta}}-1)e^{iq(x+y)}\right) & , \, x<0 & , \, y<0 \end{array}\right.  .
\ee

\be
T_\omega^{(\alpha,s)}(x,y)=\left\{\begin{array}{ccc} \left(\alpha +\frac{i\alpha^2}{\Delta_{\rm s\delta}} \right) \delta(x)\delta(y)& , \, x>0 & , \, y>0 \\ \\ \frac{i\alpha^2}{\Delta_{\rm s\delta}}  \delta(x)\delta(y)+\frac{i\alpha s^2}{\Delta_{\rm s\delta}} e^{-iq y}\delta(x)& , \, x>0 & , \, y<0 \\ \\ \frac{i\alpha^2}{\Delta_{\rm s\delta}}  \delta(x)\delta(y)+\frac{i\alpha s^2}{\Delta_{\rm s\delta}} e^{-iq x}\delta(y) & , \, x<0 & , \, y>0 \\ \\ \left(\alpha+\frac{i\alpha^2}{\Delta_{\rm s\delta}} \right) \delta(x)\delta(y)+\frac{i\alpha s^2}{\Delta_{\rm s\delta}} \left(e^{-iq y}\delta(x)+e^{-iq x}\delta(y)\right)  & , \, x<0 & , \, y<0 \end{array}\right.  .
\ee
\end{widetext}

From these formulas for the Green's function and the kernel of the ${\bf T}$-matrix one derives the kernel of the two step/$\delta$'s ${\bf M}$-matrix by shifting the origin to the $x=\pm a$ points and bearing in mind that at $x=a$ one must exchange $k$ and $p$ in the solutions at the left and the right of the ${\rm s-\delta}$-potential.

\subsection{Miscellanea: some exotic backgrounds}

Other more exotic backgrounds include two linear combinations of $\delta$ and $\delta^\prime$ interactions
concentrated at the points $x=\pm a$:
\ben
U(x)&=&\alpha_1 \delta(x+a)+\lambda_1 \delta^\prime (x+a)\nonumber\\&+&\alpha_2 \delta(x-a)+\lambda_2 \delta^\prime (x-a) \label{ddprime} \, \, .
\een
We understand the $\delta^\prime$ interaction in (\ref{ddprime}) as the self-adjoint extension of the free particle Hamiltonian proposed in reference \cite{Gadella20091310}. Because this potential responds to interactions concentrated in isolated points the general expression (\ref{genT-pointV}) provides us with the kernel of the ${\bf T}$-matrix for a $\delta/\delta^\prime$ interaction at the origin:
\be
T_\omega^{(\alpha , \lambda)}(x_1,x_2)=\frac{2\kappa(\alpha+\kappa\lambda^2)}{2\kappa(1+\frac{\lambda^2}{4})+\alpha}\delta(x_1)\delta(x_2) \nonumber \, .
\ee
The reduced Green's function is, of course, the free particle Green's function and simply application
of the $TGTG$ formula, after the appropriate shiftings to $x=\pm a$ and use of these ingredients, leads to the Casimir energy. We skip writing the quantum vacuum energy of the double $\delta/\delta^\prime$ system but we mention that re-interpretation of these point interactions as boundary conditions encompass for some special values of the parameters not only Dirichlet but also Neumann and even Robin boundary conditions.

A promising avenue is the study of scalar field fluctuations on solitonic or curved back grounds. For instance, the background
\ben
U(x)&=&\alpha \delta(x+a)+\beta \delta(x-a) +\nonumber \\ &+& m^2\left(1- \theta(a-x)\theta(a+x)\cdot\frac{2}{\cosh^2 m x}\right) \nonumber \, ,
\een
see \cite{CaveroPelaez:2009vi}, can be interpreted in comparison with the Hamiltonian ${\bf K}^{\rm sG}$
(\ref{sghes}) as a sine-Gordon kink background constrained to the finite interval $(-a,a)$ with two $\delta$-function interactions at the endpoints. In this paper the Green's function and the energy-momentum
tensor of the double delta/P$\ddot{\rm o}$sch-Teller configuration were computed paving the way to the calculation of the quantum vacuum energy. In reference \cite{Guilarte:2010xn} two of us analysed the scattering problem for this potential. In particular, the scattering amplitudes, as well as the bound state structure, were fully unveiled. The limit of Dirichlet boundary conditions revealed many subtleties, particularly relevant to the ground state (a zero mode) of the system. By switching on the $\delta^\prime$-interactions on the endpoints more general boundary conditions can be achieved. Nevertheless, the full analysis of the quantum vacuum energy remains to be performed. More interestingly,
replacing $G_\omega^{(0)}$ by $G_\omega^{({\rm PT})}$, the particle Green's function in the P$\ddot{\rm o}$sch-Teller background, we should be able to generalise the $TG$ procedure in this situation.

The last background that we consider is formed by two $\delta$-interactions placed on antipodal
points on a circle around the origin in the $\mathbb{R}^2$-plane. If $\vec{e}_a$, $a=1,2$, $\vec{e}_a\cdot\vec{e}_b=\delta_{ab}$, is an orthonormal basis in this plane, $\vec{x}=x_1\vec{e}_1+x_2\vec{e}_2$ denotes the particle position and $\vec{a}=a_1\vec{e}_1+a_2\vec{e}_2$
is a fixed vector, we write the background and the 2D Schr$\ddot{\rm o}$dinger operator that governs the fluctuations of the scalar field in two spatial dimensions in the form:
\begin{equation*}
U(\vec{x})=\alpha_- \delta^{(2)}(\vec{x}+\vec{a})+\alpha_+ \delta^{(2)}(\vec{x}-\vec{a}) 
\end{equation*}
\begin{equation*}
 K^{2\delta}= -\nabla^2+m^2+U(\vec{x})=-\frac{\partial^2}{\partial x_1^2}-\frac{\partial^2}{\partial x_2^2}+m^2+U(\vec{x}) \, ,
\end{equation*}
where $\delta^{(2)}(\vec{v})=\delta(v_1)\delta(v_2)$.

Fourier analysis unveils the scattering wave solutions of the spectral problem $K^{2\delta}\psi_\omega(\vec{x})=\omega^2\psi_\omega(\vec{x})$:
\ben
\psi_\omega(\vec{x})&=&e^{i\vec{k}\vec{x}}
 - \int\frac{d^2p}{(2\pi)^2}\frac{e^{-i \vec{p}\vec{x}}}{\vert\vec{p}\vert^2-\vert\vec{k}\vert^2-i \varepsilon}\times\nonumber \\ &&\times\left\{\alpha_-e^{-i \vec{p}\vec{a}}\psi_\omega(-\vec{a})+\alpha_+e^{i \vec{p}\vec{a}}\psi_\omega(\vec{a})\right\}\nonumber \, ,
\een
where $\vert\vec{k}\vert^2=\omega^2-m^2$ and the scattering amplitudes are:
\ben
&& \psi_\omega(\pm\vec{a})=\frac{1+\alpha_\mp\left(I_1[\vert\vec{k}\vert]-I_2[\vert\vec{k}\vert ,\vert\vec{a}\vert]\right)}{\Delta (\vert\vec{k}\vert ,\vert\vec{a}\vert , \alpha_\pm)}\nonumber \\
&& I_1[\vert\vec{k}\vert]=\int\frac{d^2p}{(2\pi)^2}\frac{1}{\vert\vec{p}\vert^2-\vert\vec{k}\vert^2-i \varepsilon}\nonumber
\een
\ben
 && I_2[\vert\vec{k}\vert ,\vert\vec{a}\vert]= \int\frac{d^2p}{(2\pi)^2}\frac{e^{\pm 2 i \vec{p}\vec{a}}}{\vert\vec{p}\vert^2-\vert\vec{k}\vert^2-i \varepsilon} \nonumber 
\een
\ben 
 && \Delta (\vert\vec{k}\vert ,\vert\vec{a}\vert , \alpha_\pm)=\left(1+\alpha_+I_1[\vert\vec{k}\vert] \right)\left(1+\alpha_-I_1[\vert\vec{k}\vert]\right)-\nonumber \\ && \hspace{2.3cm}-\alpha_+\alpha_- I_2^2[\vert\vec{k}\vert ,\vert\vec{a}\vert]\nonumber \, \, .
\een
The integrals $I_1$ and $I_2$ are ultraviolet divergent but can be regularised by using, e.g., a cutoff in the momentum. The identification of the scattering data from these solutions is extremely delicate from an analytical point of view because the lack of rotational invariance of the potential.

Fortunately, the $TG$ formalism only requires addressing the problem of one centre (and the knowledge of the free particle Green's function), an issue solved by Jackiw in \cite{Jackiw:1991je} where he found two striking results: (1) the scale invariance of the
$\delta^{(2)}$-interaction is broken by the quatization process; (2) The scattering waves only emerge
in $s$-waves, the only non null phase shifts corresponding to zero angular momentum. To work Jackiw formulas
in a point away from the origin is the remaining task to perform in order to compute the two $\delta^{(2)}$-lines Casimir energy from the $TGTG$ formula. Physically this mathematical structure arises as quantized vortex lines in superfuid Helium 4 or other Bose condensates. If the phase of the order parameter (the complex scalar field) is of the form $\chi_v(\vec{x},t)=
\frac{l}{\pi}, {\rm arctan}\frac{x_2}{x_1}$, $l\in\mathbb{Z}$, the gradient of the fluid velocity $\vec{\nabla}\chi_v(\vec{x},t)=\frac{l}{\pi}\left(\frac{x_1}{\vert\vec{x}\vert^2}\vec{e}_1+\frac{x_2}{\vert\vec{x}\vert^2}\vec{e}_2\right)$ is precisely: $\nabla^2\chi_v(\vec{x},t)=\frac{l}{\pi} \delta^{(2)}(\vec{x})$. Therefore, the $TGTG$ formalism allows the calculation of the quantum vacuum energies between two vortex lines in superfluid liquids  (see ref. \cite{donnelly1991quantized}).
\section*{Acknowledgements}
 The authors would like to thank Michael Bordag, I. Cavero-Pelaez, M. Asorey and G. Marmo for illuminating discussions, and S. Ratcliffe for language support. This work has been supported by the German DFG grant BO 1112/18-1, and the European Union ESF Research Network CASIMIR.

\bibliography{2delta-bibliogr}
\bibliographystyle{unsrt}



\end{document}